\documentclass[
	prl,
	reprint,
	article,
	twocolumn,
	nofootinbib,
	floatfix,
	superscriptaddress
]{revtex4-2}

\usepackage{graphicx}% Include figure files
\usepackage{dcolumn}% Align table columns on decimal point
\usepackage{bm}% bold math

\usepackage[utf8]{inputenc}
\usepackage[T1]{fontenc}
\usepackage{mathptmx}
\usepackage{amssymb,amsmath}
\usepackage{amsfonts}
\usepackage{hyperref}

\usepackage[dvipsnames]{xcolor}

\newcommand{\kb}{k_{\text{B}}} % Boltzmann constant

\begin{document}

\title[Active Refrigerators Powered by Inertia]{Active Refrigerators Powered by Inertia}

\author{Lukas~Hecht}
 \affiliation{ 
 	Institut f\"ur Physik kondensierter Materie, Technische Universit\"at Darmstadt, Hochschulstr. 8, D-64289 Darmstadt, Germany
 }
\author{Suvendu~Mandal}
 \affiliation{ 
	Institut f\"ur Physik kondensierter Materie, Technische Universit\"at Darmstadt, Hochschulstr. 8, D-64289 Darmstadt, Germany
}
\author{Hartmut~L\"owen}
\affiliation{ 
	Institut f\"ur Theoretische Physik II - Soft Matter, Heinrich-Heine-Universit\"at D\"usseldorf, Universit\"atsstraße 1, D-40225 D\"usseldorf, Germany
}
\author{Benno~Liebchen}
\email{benno.liebchen@pkm.tu-darmstadt.de}
\affiliation{ 
	Institut f\"ur Physik kondensierter Materie, Technische Universit\"at Darmstadt, Hochschulstr. 8, D-64289 Darmstadt, Germany
}

\date{\today}

\begin{abstract}
We present the operational principle for a refrigerator which uses inertial effects in active Brownian particles to locally reduce their (kinetic) temperature by two orders of magnitude below the environmental temperature. This principle exploits the peculiar but so-far unknown shape of the phase diagram of inertial active Brownian particles to initiate motility-induced phase separation in the targeted cooling regime only. Remarkably, active refrigerators operate without requiring isolating walls opening the route towards using them to systematically absorb and trap, e.g., toxic substances from the environment.
\end{abstract}

\maketitle

\begin{figure}
	\centering
	\includegraphics[width=1.0\linewidth]{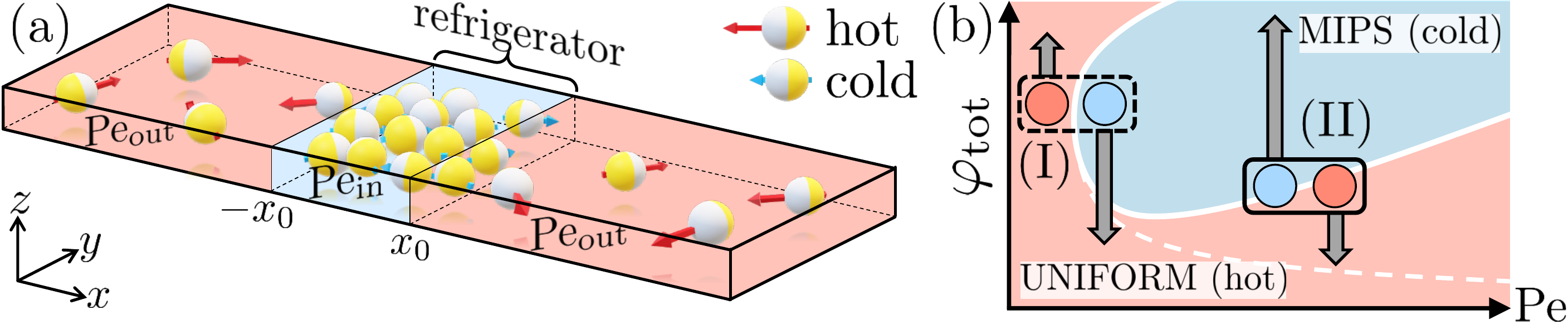}
	\caption{Schematic of the active refrigerator (a), which exploits the peculiar shape of the phase diagram (b). The blue region represents phase coexistence (MIPS), the white solid line the newly discovered transition line for inertial ABPs in comparison with the well-known transition line for overdamped ABPs (dashed line). Boxes and arrows refer to relevant parameter regimes discussed in the text.}
	\label{fig:fig1}
\end{figure}

\begin{figure}
	\centering
	\includegraphics[width=1.0\linewidth]{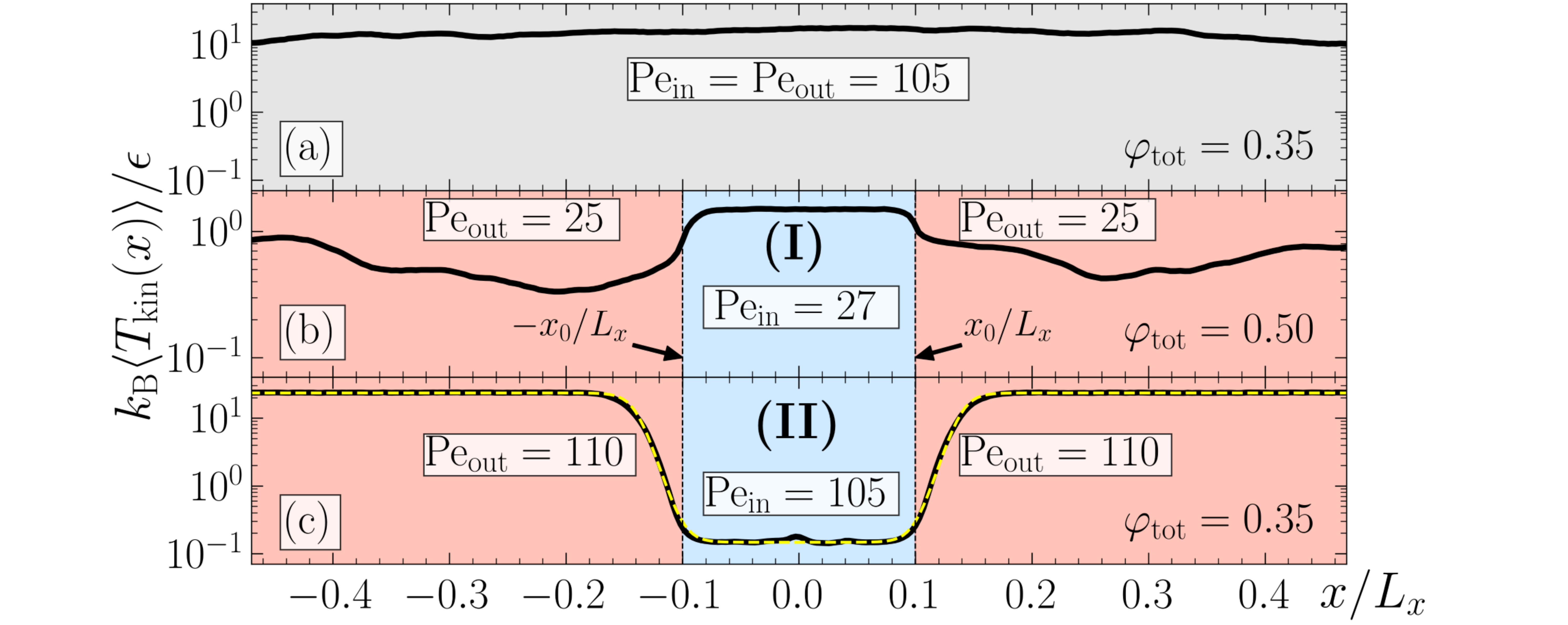}
	\caption{Kinetic temperature profiles $k_{\rm B}T_{\rm kin}(x)=m\langle |\vec{v}|^2\rangle_y/2$ in the steady state averaged over the $y$ coordinate and 20 realizations with $N=16000$ particles for uniform Pe (a) and nonuniform Pe (b)--(c) and parameters shown in the key. The yellow dashed line is a fit of $f(x)=a(2-\tanh(b(x+c))+\tanh(b(x-c)))/2+d$.}
	\label{fig:fig2}
\end{figure}

\textit{Introduction} ---
Many processes in nature allow to readily heat up an isolated system. Examples include the release of heat in chemical reactions occurring, e.g., when burning wood or gas, inelastic collisions occurring within resistors when exposed to electric currents, and mass-energy conversion processes in nuclear power plants and helium-burning stars. Following the second law of thermodynamics, none of these processes can be reverted, making us believe that it is impossible to cool down an isolated physical system. Accordingly, cooling down a target domain such as the inside of a refrigerator or atoms in a magneto-optical trap requires that the relevant domain is in contact with an external bath to which heat can be transferred via conduction, convection, radiation, or evaporation. Accordingly, developing sophisticated techniques to transfer heat from a target system to the environment has been a great challenge of twenties century physics \cite{Metcalf_Book_LaserCoolingAndTrapping_1999,Schroeder_Book_AnIntroductionToThermalPhysics_2000,Ziabari_RepProgPhys_2016,Letokhov_QuantSemiclassOpt_1995,VanDenBroeck_PhysRevLett_2006}.
\\For active systems \cite{Bechinger_RevModPhys_2016,Cates_AnnuRevCondensMatterPhys_2015,Marchetti_RevModPhys_2013,Elgeti_RepProgPhys_2015,Gompper_JPhysCondensMatter_2020,Bowick_PhysRevX_2022,Liebchen_JPhysCondensMatter_2022,Schildknecht_SoftMatter_2022}, which consist of self-propelled particles and are intrinsically out of equilibrium, the second law does not apply to the active particles (but only to the overall system) \cite{SI}. Therefore, in the present work, we ask if it is possible to cool down a system of active Brownian particles (ABPs) \cite{Romanczuk_EurPhysJSpecialTopics_2012,Bechinger_RevModPhys_2016} \emph{in a certain target region} [refrigerator, Fig.\ \ref{fig:fig1}(a)] in terms of their kinetic temperature \cite{SI} without requiring a mechanism to transfer energy to particles in the (spatially separated) environment. 
\\To achieve this, we exploit the previous finding that ABPs can spontaneously phase separate into a dense and a dilute phase (motility-induced phase separation; MIPS) \cite{Cates_AnnuRevCondensMatterPhys_2015,Tailleur_PhysRevLett_2008,Fily_PhysRevLett_2012,Cates_EPL_2013,Buttinoni_PhysRevLett_2013,Stenhammar_PhysRevLett_2013,Redner_PhysRevLett_2013,Buttinoni_PhysRevLett_2013,Stenhammar_SoftMatter_2014,Mokhtari_EPL_2017,Patch_PhysRevE_2017,Levis_SoftMatter_2017,Siebert_PhysRevE_2018,Solon_NewJPhys_2018,Durve_EPJE_2018,Digregorio_PhysRevLett_2018,Patch_SoftMatter_2018,Loewen_JCP_2020,Dai_SoftMatter_2020,Chennakesavalu_JCP_2021,Su_NewJPhys_2021,Turci_PhysRevLett_2021}. While MIPS behaves similarly to an equilibrium phase transition at large scales in the overdamped limit \cite{Levis_SoftMatter_2017,Redner_PhysRevLett_2013,Turci_PhysRevLett_2021,OByrne_PhysRevLett_2020}, in the presence of inertia, as relevant for, e.g., activated dusty plasmas \cite{Morfill_RevModPhys_2009,Nosenko_PhysRevRes_2020} or vibrating granular particles \cite{Scholz_NatCom_2018,Scholz_NewJPhys_2016,Kudrolli_PhysRevLett_2008,Weber_PhysRveLett_2013,Walsh_SoftMatter_2017,Patterson_PhysRevLett_2017,Deblais_PhysRevLett_2018,Giomi_PrcRSocA_2013,Dauchot_PhysRveLett_2019,Deseigne_PhysRevLett_2010}, the coexisting phases feature different temperatures, which is, in contrast to clustering in granular gases caused by inelastic collisions \cite{Goldhirsch_PhysRevLett_1993,Paolotti_PhysRevE_2004,Garzo_PhysRevE_2018,Fullmer_AnnRevFluidMech_2017,Puglisi_PhysRevLett_1998}, a consequence of self propulsion and elastic collisions \cite{Mandal_PhysRevLett_2019,Petrelli_PhysRevE_2020}. However, this finding alone is not sufficient to design an active refrigerator, because it leads to a dense and cold phase which occurs as randomly distributed clusters which move, merge, and coarsen, and ultimately lead to a uniform temperature profile when averaging over many realizations or a long time [Fig.\ \ref{fig:fig2}(a)]. 
\\Thus, to create an active refrigerator, we need to meet the challenge of finding a mechanism allowing us to initiate MIPS in the targeted cooling domain only and to localize the dense phase in that region. To achieve this, one naive approach could be to implement a nonuniform motility \cite{lozano2016phototaxis,lozano2019propagating} (e.g., through controlling the laser intensity in light-fueled swimmers \cite{Heidari_Langmuir_2020,Golestanian_PhysRevLett_2012,Jiang_PhysRevLett_2010,Buttinoni_JPhysCondMat_2012}) such that particles in the targeted cooling domain show a (large) P\'eclet number (Pe; relative importance of self propulsion compared to diffusion) beyond the critical one for the MIPS phase transition, whereas particles in the environment feature a (small) sub-critical Pe [Fig.\ \ref{fig:fig1}(b), regime (I)]. However, this does not work because Pe and density essentially behave inversely to each other \cite{Stenhammar_SoftMatter_2014,Cates_AnnuRevCondensMatterPhys_2015} such that locally increasing Pe decreases the density in the same spatial region and does not result in a significant temperature difference [Fig.\ \ref{fig:fig2}(b)]. Remarkably, however, the opposite strategy turns out to work in a carefully selected portion of the phase diagram [Fig.\ \ref{fig:fig1}(b), regime (II)]: we find that reducing Pe in the targeted cooling domain by less than $5\%$ as compared to the environment reduces the kinetic temperature of the ABPs by two orders of magnitude. This surprising finding exploits a remarkable difference between the phase diagram of inertial ABPs and the well-known phase diagram of overdamped ABPs: while MIPS occurs in overdamped ABPs when both Pe and the density are sufficiently large, in underdamped ABPs, it occurs at sufficiently large density and intermediate Pe. Thus, when choosing values of Pe within this intermediate regime in the targeted cooling domain and higher values in the environment, the density further increases in the former region bringing the system deeper into the MIPS regime and further away from it outside. That is, inertia is required twice: first, to induce the two-temperature coexistence and second, to create the required shape of the phase diagram. 
\\The resulting active refrigerator exemplifies a fundamentally new way to locally cool down a physical system. Like ordinary refrigerators, it can be used to cool down other objects. However, as opposed to ordinary cooling devices, active refrigerators use a self-organized cooling domain such that no isolating walls are required to separate the cooling domain from its environment. As a consequence, active refrigerators can in principle also be used as a device to absorb particles from the environment and to store them for a long time, as we shall see.

%%%%%%%%%%%%%%%%%%%%%%%%%%%%%%%%%%%%%%%%%%%%%%%%%%%%%%%%%%%%%%%%%%%%%%%%%%%%%%%%%%%%%%%%%%%%%%%%%%%%%%%%%%%%%
% SIMULATION MODEL
%%%%%%%%%%%%%%%%%%%%%%%%%%%%%%%%%%%%%%%%%%%%%%%%%%%%%%%%%%%%%%%%%%%%%%%%%%%%%%%%%%%%%%%%%%%%%%%%%%%%%%%%%%%%%
\textit{Model} --- 
We consider inertial active Brownian particles (ABPs) \cite{Romanczuk_EurPhysJSpecialTopics_2012,Mandal_PhysRevLett_2019,Loewen_JCP_2020,Gutierrez-Martinez_JCP_2020,Sandoval_PhysRevE_2020,Su_NewJPhys_2021} in two spatial dimensions. Each particle is represented by a (slightly soft) disk of diameter $\sigma$, mass $m$, and moment of inertia $I=m\sigma^2/10$ and features an effective self-propulsion force $\vec{F}_{\text{SP},i}=\gamma_{\text{t}}v_0\hat{p}_i(t)$, where $v_0,\hat p_i$ denote the (terminal) self-propulsion speed and the orientation $\hat{p}_i(t)=(\cos\phi_i(t),\sin\phi_i(t))$ of the $i$-th particle ($i=1,2..N$), respectively. Position $\vec{r}_i$ and orientation angle $\phi_i$ evolve according to $\text{d}\vec{r}_i/\text{d}t=\vec{v}_i$ and $\text{d}\phi_i/\text{d}t=\omega_i$, respectively, where the velocity $\vec{v}_i$ and the angular velocity $\omega_i$ in turn evolve as
\begin{align}
	m\frac{\text{d}\vec{v}_i}{\text{d} t} =&~-\gamma_{\text{t}}\vec{v}_i+\gamma_{\text{t}}v_0\hat{p}_i - \sum_{\substack{j=1\\j\neq i}}^{N}\nabla_{\vec{r}_i}u\left(r_{ij}\right) + \sqrt{2\kb T_{\text{b}}\gamma_{\text{t}}}\vec{\xi}_i,\\ 
	I\frac{\text{d}\omega_i}{\text{d} t} =&-\gamma_{\text{r}}\omega_i+\sqrt{2\kb T_{\text{b}}\gamma_{\text{r}}}\eta_{i}.
\end{align}
Here, $\gamma_{\text{t}}$, $\gamma_{\text{r}}$ are the translational and rotational drag coefficients, respectively, and $T_{\text{b}}$ is the temperature of the bath, e.g., of the liquid/plasma medium surrounding the particles, which can differ from the kinetic temperature of the particles \cite{Falasco_PhysRevE_2014} and which we treat as constant in our simulations (see Supplemental Material (SM) \cite{SI}). The interaction potential $u(r_{ij})$, $r_{ij}=\left|\vec{r}_i-\vec{r}_j\right|$ is modeled by the Weeks-Chandler-Anderson (WCA) potential \cite{Weeks_JCP_1971} with strength $\epsilon$ and effective particle diameter $\sigma$. Finally, $\vec{\xi}_i$ and $\eta_{i}$ denote Gaussian white noise with zero mean and unit variance. We define $\text{Pe}=v_0/\sqrt{2D_{\text{r}}D_{\text{t}}}$, where $D_{\text{t}}=\kb T_{\text{b}}/\gamma_{\text{t}}$ and $D_{\text{r}}=\kb T_{\text{b}}/\gamma_{\text{r}}$ denote the translational and rotational diffusion coefficients, respectively. Note that ABP models like ours do not explicitly describe the self-propulsion mechanism, the underlying energy source or how energy is dissipated into the bath \cite{Romanczuk_EurPhysJSpecialTopics_2012,Hecht_ArXiv_2021}. We discuss possible experimental realizations below and develop a thermodynamically consistent picture in the paragraph ``where does the energy flow?''.
\\In all simulations, we fix $m/(\gamma_{\text{t}}\tau_{\rm p})=5\times10^{-2}$, $I/(\gamma_{\text{r}}\tau_{\rm p})=5\times10^{-3}$, $\epsilon/(k_{\text{B}}T_{\text{b}})=10$, and $\sigma/\sqrt{D_{\rm r}D_{\rm t}}=1$ with the persistence time $\tau_{\rm p}=1/D_{\rm r}$. We choose $\gamma_{\text{t}}=\gamma_{\text{r}}/\sigma^2$ and vary Pe and the total area fraction $\varphi_{\text{tot}}=N\pi\sigma^2/(4A)$, where $A=L_xL_y,~L_y/L_x=0.05$, denotes the area of the simulation box. The Langevin equations are solved numerically with \verb|LAMMPS| \cite{Plimpton_JCompPhys_1995,Thompson_CompPhysComm_2022} for up to $N=10^{5}$ particles using periodic boundary conditions and a time step $\Delta t/\tau_{\rm p}=10^{-5}$ (see SM \cite{SI} for further details).
\\Our setup is illustrated in Fig.\ \ref{fig:fig1}(a): the simulation area is divided into two regions, in which the particles have different P\'eclet numbers $\text{Pe}(x_i)=v_0(x_i)/\sqrt{2D_{\text{r}}D_{\text{t}}}$, i.e., the self-propulsion speed of each particle depends on its position according to
\begin{equation}
	v_0(x_i)=\begin{cases}
	v_{0,\text{in}},~-x_0<x_i<x_0\\
	v_{0,\text{out}},~\text{else}
	\end{cases},
	\label{eq:v_of_x}
\end{equation}
with $x_0\ll L_x$. Note that our results are robust with respect to changes of $x_0$, $N$, $m$, $v_{0,\text{in}}$, and $v_{0,\text{out}}$ and in particular, apply to values of $m/(\gamma_{\text{t}}\tau_{\rm p})$ used in previous works \cite{Petrelli_PhysRevE_2020,Su_NewJPhys_2021,Caprini_SoftMatter_2021,Takatori_PhysRevFluids_2017,Dai_SoftMatter_2020,Mandal_PhysRevLett_2019} (Figs.\ S9--S12 in the SM \cite{SI}). Initially, all particles are uniformly distributed in the whole simulation area.

%%%%%%%%%%%%%%%%%%%%%%%%%%%%%%%%%%%%%%%%%%%%%%%%%%%%%%%%%%%%%%%%%%%%%%%%%%%%%%%%%%%%%%%%%%%%%%%%%%%%%%%%%%%%%
% SUMMARY OF MAIN RESULTS
%%%%%%%%%%%%%%%%%%%%%%%%%%%%%%%%%%%%%%%%%%%%%%%%%%%%%%%%%%%%%%%%%%%%%%%%%%%%%%%%%%%%%%%%%%%%%%%%%%%%%%%%%%%%%

\textit{Active refrigerators} --- 
The goal is now to find ${\rm Pe}_{\text{in}}$ and ${\rm Pe}_{\text{out}}$ such that (i) MIPS occurs in the targeted cooling domain only and (ii) the resulting dense phase stays in that region. Notice first, that when choosing $\text{Pe}_{\text{in}}=\text{Pe}_{\text{out}}$, in each individual realization, we find different kinetic temperatures in coexisting phases, but the ensemble-averaged (or time-averaged) kinetic temperature profile is uniform [Fig.\ \ref{fig:fig2}(a)]. If we choose $\varphi_{\rm tot}=0.5$ and ${\rm Pe}_{\text{in}}>{\rm Pe}_{\text{out}}$ [regime (I) in Fig.\ \ref{fig:fig1}(b)] to trigger MIPS in the target domain only, however, we obtain only a weak temperature difference (which even goes in the wrong direction), because the particle density compensates the difference in Pe (because the residential time of a particle in a small volume element scales inversely to its speed) as indicated by the gray arrows in Fig.\ \ref{fig:fig1}(b) (note that the arrow length depends on the density of both phases and thus, is not obvious). More generally, when choosing other combinations ${\rm Pe}_{\text{in}}>{\rm Pe}_{\text{out}}$ and density in the left part of the phase diagram [Fig.\ \ref{fig:fig1}(b), regime (I)], we do not observe any relevant cooling in the target domain. Remarkably, however, if we choose a comparatively low area fraction of $\varphi_{\rm tot}=0.35$ and ${\rm Pe}_{\text{in}}=105<{\rm Pe}_{\text{out}}=110$ [regime (II) in Fig.\ \ref{fig:fig1}(b)], we observe that the system undergoes MIPS exclusively in the target domain and the dense phase remains in that region (Movie M1 in the SM \cite{SI}). This results in a striking cooling effect by more than two orders of magnitude in the cooling domain from $\kb \langle T_{\rm kin}^{\rm (out)}\rangle/\epsilon\approx 23.4$ to $\kb \langle T_{\rm kin}^{\rm (in)}\rangle/\epsilon\approx 0.147$ [Fig.\ \ref{fig:fig2}(c)], which is further enhanced when choosing larger Pe differences and complemented by a significantly lower entropy production rate in the cooling domain and an inward flow of kinetic energy (Figs.\ S3--S5 in the SM \cite{SI}).

%%%%%%%%%%%%%%%%%%%%%%%%%%%%%%%%%%%%%%%%%%%%%%%%%%%%%%%%%%%%%%%%%%%%%%%%%%%%%%%%%%%%%%%%%%%%%%%%%%%%%%%%%%%%%
% RESULTS AND EXPLANATIONS
%%%%%%%%%%%%%%%%%%%%%%%%%%%%%%%%%%%%%%%%%%%%%%%%%%%%%%%%%%%%%%%%%%%%%%%%%%%%%%%%%%%%%%%%%%%%%%%%%%%%%%%%%%%%%
\begin{figure}
	\centering
	\includegraphics[width=1.0\linewidth]{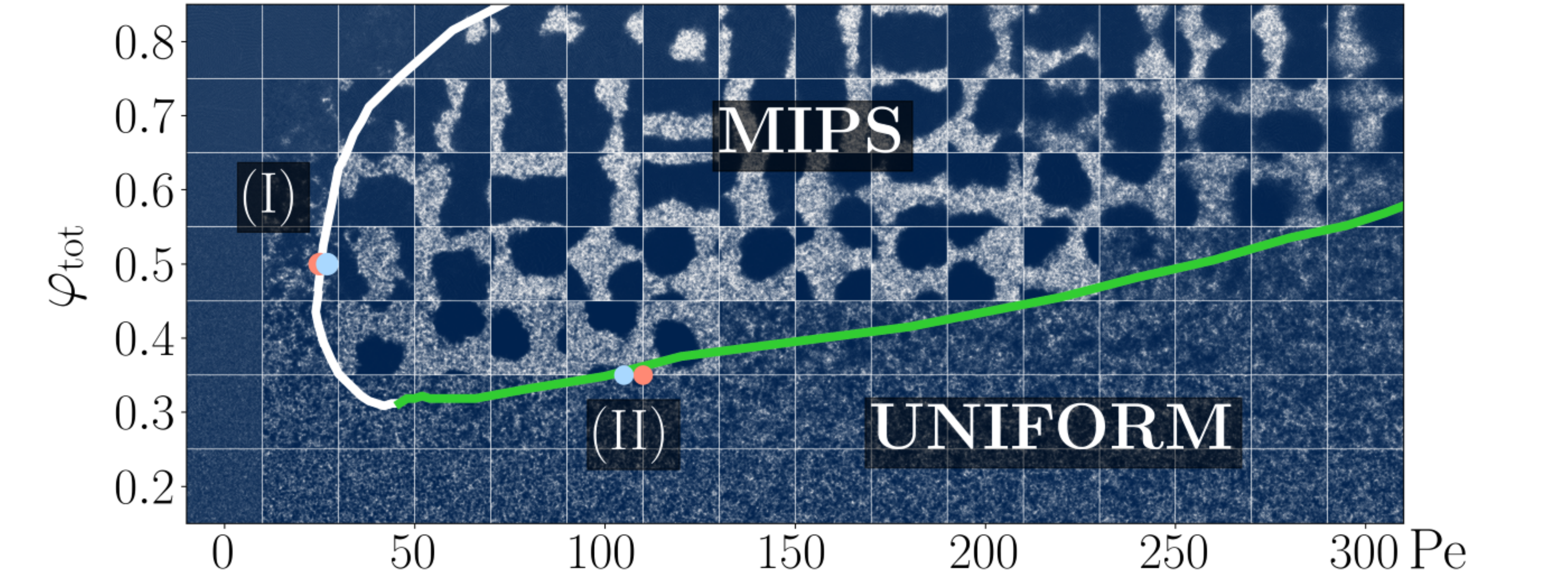}
	\caption{Phase diagram of $N=20000$ inertial ABPs (background images are steady-state snapshots). The solid line shows the transition line (see SM \cite{SI} for details). In the vicinity of its green part, parameters can be chosen to construct active refrigerators.}
	\label{fig:fig3}
\end{figure}

\textit{Phase diagram} --- 
To understand the possible parameter choices for constructing active refrigerators in detail, we now discuss the phase diagram of inertial ABPs in the Pe-$\varphi_{\rm tot}$-plane, which has remained unknown to date. The key control parameters of the system are $\epsilon$, Pe, and $\varphi_{\rm tot}$ for fixed $m$ and $I$. We additionally fix $\epsilon$ and vary Pe and $\varphi_{\rm tot}$. To determine the transition line between the uniform state and the MIPS regime (Fig.\ \ref{fig:fig3}), we investigate the distribution of the local area fraction $\varphi_{\rm loc}$ \cite{Digregorio_PhysRevLett_2018,Su_NewJPhys_2021,Klamser_NatCom_2018}, which is unimodal in the uniform regime and bimodal in the coexistence regime (Fig.\ S1 in the SM \cite{SI}). Interestingly, the transition line does not follow the well-known relation ${\rm Pe}\propto 1/\varphi_{\text{tot}}$, which was found in the overdamped regime \cite{Tailleur_PhysRevLett_2008,Cates_EPL_2013}. In striking contrast, we find that ${\rm Pe}\propto\varphi_{\text{tot}}$ in the large Pe regime (green part of the transition line in Fig.\ \ref{fig:fig3}). This relation serves as a crucial ingredient to construct an active refrigerator. Intuitively, it can be understood to occur as a direct consequence of inertial effects: the particles bounce back when they collide with each other and the rebound is much stronger for large Pe than for moderate Pe. Therefore, to slow down locally, more collisions are necessary and a larger area fraction is required at larger Pe to initiate MIPS.

\textit{Design rule} ---
Based on the transition line, we can formulate the following strategy to realize the active refrigerator: first, we want to initiate MIPS in the target domain. This can be achieved by choosing $({\rm Pe}_{\rm in}, \varphi_{\rm in})$ inside the MIPS region of the phase diagram for the target domain. Second, we do not want the system to undergo MIPS outside the target domain. Hence, we choose $({\rm Pe}_{\rm out}, \varphi_{\rm out})$ outside the coexistence region. Third, we want the particle flux which emerges as a consequence of choosing two different Pe to bring the system deeper into the coexistence regime within the target domain but further away from it outside. Clearly, based on the obtained detailed knowledge of the phase transition line, the first two criteria can be easily met by fixing a suitable area fraction $\varphi_{\rm in}=\varphi_{\rm out}=\varphi_{\rm tot}$ and choosing two P\'eclet numbers on both sides of the transition line. However, the third criterion can only be met by choosing parameter combinations in the vicinity of the green marked part of the transition line [regime (II)]. To see this, we will next discuss the particle flux which emerges when choosing two different Pe.

\begin{figure}
	\centering
	\includegraphics[width=1.0\linewidth]{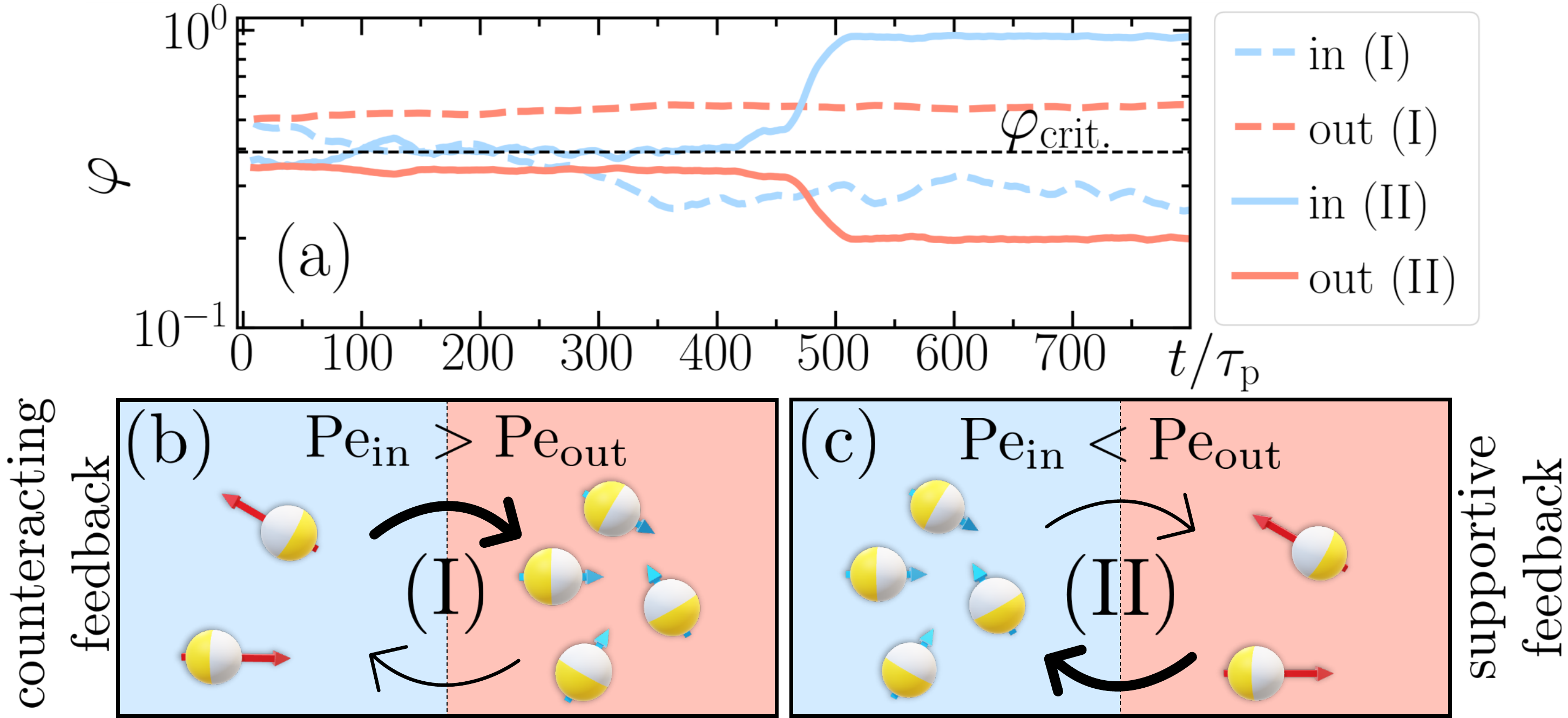}
	\caption{(a) Area fraction in inner and outer regions over time for regime (I) and (II) (parameters as in Fig.\ \ref{fig:fig2}). The dashed horizontal line shows the critical area fraction $\varphi_{\rm crit.}\approx 0.39$ for ${\rm Pe}=27$. A (b) counteracting [(c) supportive] feedback loop decreases [increases] the particle density in the target region.}
	\label{fig:fig4}
\end{figure}

\textit{Supportive and counteracting feedback} --- 
Let us first recall that the mean speed of an ABP decreases with increasing $\varphi_{\rm tot}$ and increases with increasing Pe (Fig.\ S2 in the SM \cite{SI}). Consequently, when we have two regions with different Pe, a lower density will emerge in the high-Pe region and a larger one in the low-Pe region. Therefore, the gray arrows in Fig.\ \ref{fig:fig1}(b) always point to lower $\varphi_{\rm tot}$ at the high-Pe point and vice versa.
\\In regime (I) and more generally, in the vicinity of the white part of the transition line in Fig.\ \ref{fig:fig3}, we need to choose ${\rm Pe}_{\rm in}>{\rm Pe}_{\rm critical}>{\rm Pe}_{\rm out}$ to initiate MIPS in the target domain only. Consequently, the density initially decreases in that region [Fig.\ \ref{fig:fig4}(a)]. Interestingly, the area fraction in the target domain typically decreases to values below the transition line even for a relatively small Pe difference, which fully prevents MIPS in the target domain. This surprisingly strong decrease can be viewed as the result of a positive feedback loop: the decrease of the particle density in the target domain increases the mean speed of the particles in that region, which further decreases the particle density in the target domain. Thus, no cooling occurs within that region (but rather the opposite, see Fig.\ \ref{fig:fig2}). In stark contrast, following the peculiar shape of the phase transition line at large Pe (Fig.\ \ref{fig:fig3}), the initial particle flux points into the right direction and gives rise to the enormous cooling effect for only tiny differences in Pe. More specifically, when choosing ${\rm Pe}_{\rm in}<{\rm Pe}_{\rm critical}<{\rm Pe}_{\rm out}$ [as in regime (II)], the particles are initially faster in the environment, which enhances the density inside the target domain where MIPS occurs and further slows down the particles, which further supports the particle flux from the environment.

\begin{figure}
    \centering
    \includegraphics[width=1.0\linewidth]{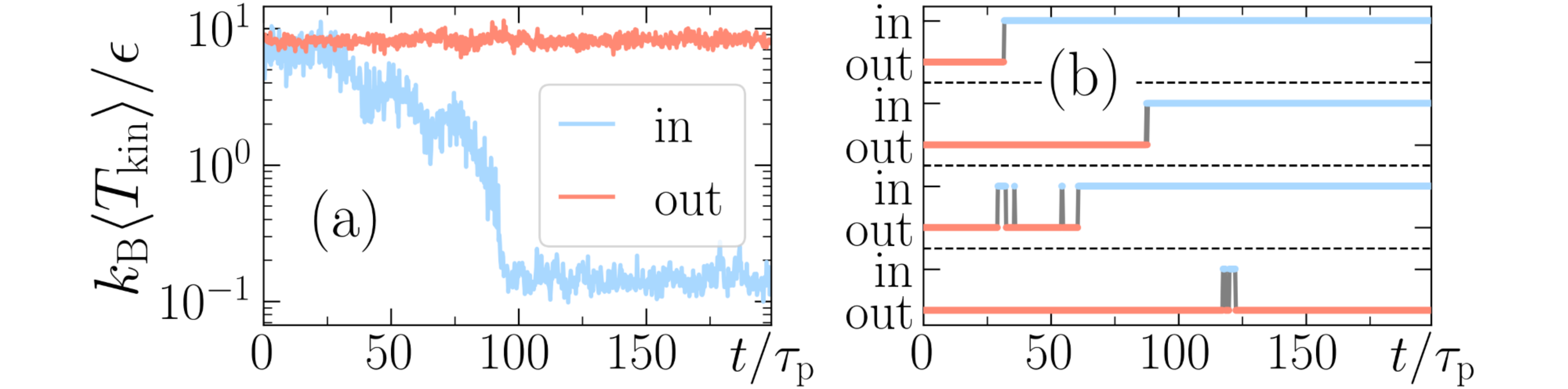}
    \caption{Absorbing, trapping, and cooling tracers with active refrigerators. (a) Kinetic temperature of passive tracers inside and outside the cooling domain. (b) Position (inside or outside the cooling domain) of four exemplary passive tracers over time [parameters as in Fig.\ \ref{fig:fig2}(c) but with ${\rm Pe}_{\rm in}={\rm Pe}_{\rm out}=0$ for passive tracers and $N_{\rm passive}/N=0.02$].}
    \label{fig:fig5}
\end{figure}

\textit{Where does the energy flow?} ---
The finding of a persistent temperature gradient for the active particles is measurable with a suitable thermometer (SM \cite{SI}) and does of course not contradict thermodynamics: heat always flows from hot to cold within the bath (solvent/gas) which surrounds the active particles. This heat flow persists in steady state and is maintained by the (external) energy source driving the system: let's imagine light-powered Janus colloids in a liquid \cite{Bechinger_RevModPhys_2016} or a complex plasma \cite{Morfill_RevModPhys_2009,Nosenko_PhysRevRes_2020}, where inertia is important. Clearly, in steady state, when neglecting temperature changes of the particle material, essentially all the energy which is absorbed by the active particles from the external light source is ultimately transferred to the bath. That is, for a uniform Pe (defocused laser), the particles act as identical heat sources for the bath. When realizing active refrigerators with a slightly nonuniform Pe (${\rm Pe}_{\rm out}\gtrsim {\rm Pe}_{\rm in}$), we obtain a significantly enhanced particle density within the refrigerator region and hence, a comparatively hot solvent. Thus, $T_{\rm b}$ is large in regions where $T_{\rm kin}$ is low, leading to a persistent bath-energy-flow from hot to cold (see SM \cite{SI} for a minimal model of $T_{\rm b}$). Note that changes in $T_{\rm b}$ are small compared to changes in $T_{\rm kin}$ since the bath has many degrees of freedom. Hence, we keep $T_{\rm b}$ constant (as typical for ABP models \cite{Bechinger_RevModPhys_2016}). (This argument is of course not restricted to light-powered swimmers but essentially applies also to, e.g., chemically powered swimmers when considering the fuel as an external energy source.)
\\The direction of the bath-energy-flow can also be spatially reverted: for ${\rm Pe}_{\rm out}\gg {\rm Pe}_{\rm in}$, the bath heats up stronger outside the refrigerator region because the light absorption grows faster than the particle density inside, which cannot exceed close packing \cite{SI}. Then, heat flows into the refrigerator region within the bath but still from hot to cold.

\textit{Absorbing, trapping, and cooling tracers with active refrigerators} ---
One unique feature of the proposed active refrigerators is that they cool down colloidal particles in a certain region in space without requiring any isolating walls separating the cooling domain from the environment. Since the kinetic temperature differences are much larger than the temperature differences in the underlying bath, active refrigerators can also be used to absorb sufficiently large substances from the environment and to trap them for a long time (Fig.\ \ref{fig:fig5}). To demonstrate this, we have performed simulations of inertial ABPs [parameters as in Fig.\ \ref{fig:fig2}(c)] and additional passive tracer particles, which may represent, e.g., certain toxic substances and are randomly distributed outside the cooling domain. Remarkably, the active refrigerator systematically absorbs tracers from the environment and cools them by two orders of magnitude below the kinetic temperature of tracers outside the refrigerator domain [Fig.\ \ref{fig:fig5}(a)]. Note that it can take a long time before a tracer enters the cooling domain, but once it is deep inside this region it stays there for a very long time, as indicated by the exemplary trajectories in Fig.\ \ref{fig:fig5}(b) and Movie M2 in the SM \cite{SI}.

\textit{Possible experimental realizations} --- 
Active refrigerators can be realized with self-propelled particles featuring significant inertia and elastic collisions such as activated micro-particles in a plasma \cite{Morfill_RevModPhys_2009,Nosenko_PhysRevRes_2020}, mesoscopic propellers such as vibrated granular particles \cite{Scholz_NatCom_2018,Scholz_NewJPhys_2016,Kudrolli_PhysRevLett_2008,Weber_PhysRveLett_2013,Walsh_SoftMatter_2017,Patterson_PhysRevLett_2017,Deblais_PhysRevLett_2018,Giomi_PrcRSocA_2013,Dauchot_PhysRveLett_2019}, drones \cite{Vasarhelyi_SciRob_2018,Duarte_ProcConfLivSys_2014,Deseigne_PhysRevLett_2010}, and mini-robots \cite{Leyman_PhysRevE_2018}, and dense animal collections \cite{Klotsa_SoftMatter_2019} such as swimming whirligig beetles as recently demonstrated in Ref.\ \cite{Devereux_JRSocInterface_2021}.

%%%%%%%%%%%%%%%%%%%%%%%%%%%%%%%%%%%%%%%%%%%%%%%%%%%%%%%%%%%%%%%%%%%%%%%%%%%%%%%%%%%%%%%%%%%%%%%%%%%%%%%%%%%%%
% SUMMARY AND OUTLOOK
%%%%%%%%%%%%%%%%%%%%%%%%%%%%%%%%%%%%%%%%%%%%%%%%%%%%%%%%%%%%%%%%%%%%%%%%%%%%%%%%%%%%%%%%%%%%%%%%%%%%%%%%%%%%%
\textit{Conclusions} ---
We have proposed a mechanism for an active refrigerator, which requires inertia not only to create a temperature difference across coexisting phases but also to induce the peculiar shape of the MIPS phase transition line, which we exploit to localize the cooling domain in a predefined region of space. As their key feature, active refrigerators create a self-organized cooling domain, in which active particles feature a much lower kinetic temperature compared to their environment. As they do not require any isolating walls to separate the cooling domain from its environment, active refrigerators prove a route towards possible future applications, e.g., to trap and absorb large (toxic) molecules or viruses. Overall, we found that the active-particle subsystem alone does not behave as one might expect from the laws of thermodynamics but makes the bath pay the thermodynamic bill for a self-organized cooling domain which does not decay. This could be further explored within microscopic theories \cite{Arold_JPCM_2020,Marconi_NewJPhys_2021}.

%\bibliography{library_mod}

% ADD SUPPLEMENTAL MATERIAL TO THIS FILE
\clearpage
\newpage

\setcounter{equation}{0}
\setcounter{figure}{0}
\setcounter{table}{0}
\setcounter{page}{1}

\renewcommand{\thefigure}{S\arabic{figure}}

\onecolumngrid

\begin{center}
		\textbf{\large Supplemental Material: Active Refrigerators Powered by Inertia}\\[.4cm]
		Lukas Hecht,$^1$ Suvendu Mandal,$^1$ Hartmut Löwen,$^2$ and Benno Liebchen$^1$\\\vspace{0.2cm}
		\small $^1$\textit{Institut für Physik kondensierter Materie, Technische Universität Darmstadt, Hochschulstr. 8, D-64289 Darmstadt, Germany}\\
		\small $^2$\textit{Institut für Theoretische Physik II - Soft Matter, Heinrich-Heine-Universität Düsseldorf, Universitätsstraße 1, D-40225 Düsseldorf, Germany}
\end{center}

\twocolumngrid
\section*{Simulation Details}
The system of Langevin equations is solved numerically by using \verb|LAMMPS| \cite{Plimpton_JCompPhys_1995,Thompson_CompPhysComm_2022}. The interaction between two particles $i$ and $j$ is modeled by the purely repulsive WCA potential \cite{Weeks_JCP_1971}
\begin{equation}
	u(r_{ij})=
	\begin{cases}
		4\epsilon\left[\left(\frac{\sigma}{r_{ij}}\right)^{12}-\left(\frac{\sigma}{r_{ij}}\right)^{6}\right]+\epsilon,~&r_{ij}/\sigma\leq 2^{1/6} \\ 0, ~&\text{else}
	\end{cases},\label{eq:wca}
\end{equation}
with $r_{ij}=|\vec{r}_i-\vec{r}_j|$, particle diameter $\sigma$, and strength $\epsilon$. By using the natural units $\tau_{\rm p}=1/D_{\rm r}$ and $l_{\rm p}=v_0\tau_{\rm p}$ (persistence time and persistence length, respectively), the Langevin equations (cf.\ Eqs. (1) and (2) in the main text) can be rewritten in dimensionless form as
\begin{align}
	m^*\frac{\text{d}\vec{v}_i^*}{\text{d} t^*} =&~-\vec{v}_i^*+\hat{p}_i + \frac{1}{\text{Pe}}\vec{\xi}_i(t^*) - \frac{1}{4T_{\rm b}^*\text{Pe}^2}\sum_{\substack{j=1\\j\neq i}}^{N}\nabla_{\vec{r}_i^*}u^*\left(r_{ij}^*\right),\\
	\frac{m^*}{10}\frac{\text{d}\omega_i^*}{\text{d} t^*} =&-\omega_i^*+\sqrt{2}\eta_{i}(t^*)
\end{align}
with reduced mass $m^*=m/(\gamma_{\rm t}\tau_{\rm p})$, P\'eclet number \text{$\text{Pe}=v_0/\sqrt{2D_{\text{r}}D_{\text{t}}}$}, and reduced bath temperature $T_{\rm b}^*=\kb T_{\rm b}/\epsilon$. The dimensionless WCA potential is given by $u^*(r_{ij}^*)=u(r_{ij})/\epsilon$ and the dimensionless variables are defined by $\vec{v}_i^*=\vec{v}_i\tau_{\rm p}/l_{\rm p}$, $\vec{r}_i^*=\vec{r}_i/l_{\rm p}$, $t^*=t/\tau_{\rm p}$, and $\omega_i^*=\omega_i\tau_{\rm p}$. Here, we use the moment of inertia $I$ of a rigid sphere, i.e., $I=m\sigma^2/10$. For fixed mass $m$, particle diameter $\sigma$, interaction strength $\epsilon$, and bath temperature $T_{\rm b}$, the leftover parameters, which control the dynamics of the system, are Pe and the total area fraction $\varphi_{\rm tot}$.

\section*{Phase Diagram}
\begin{figure}
	\centering
	\includegraphics[width=0.65\linewidth]{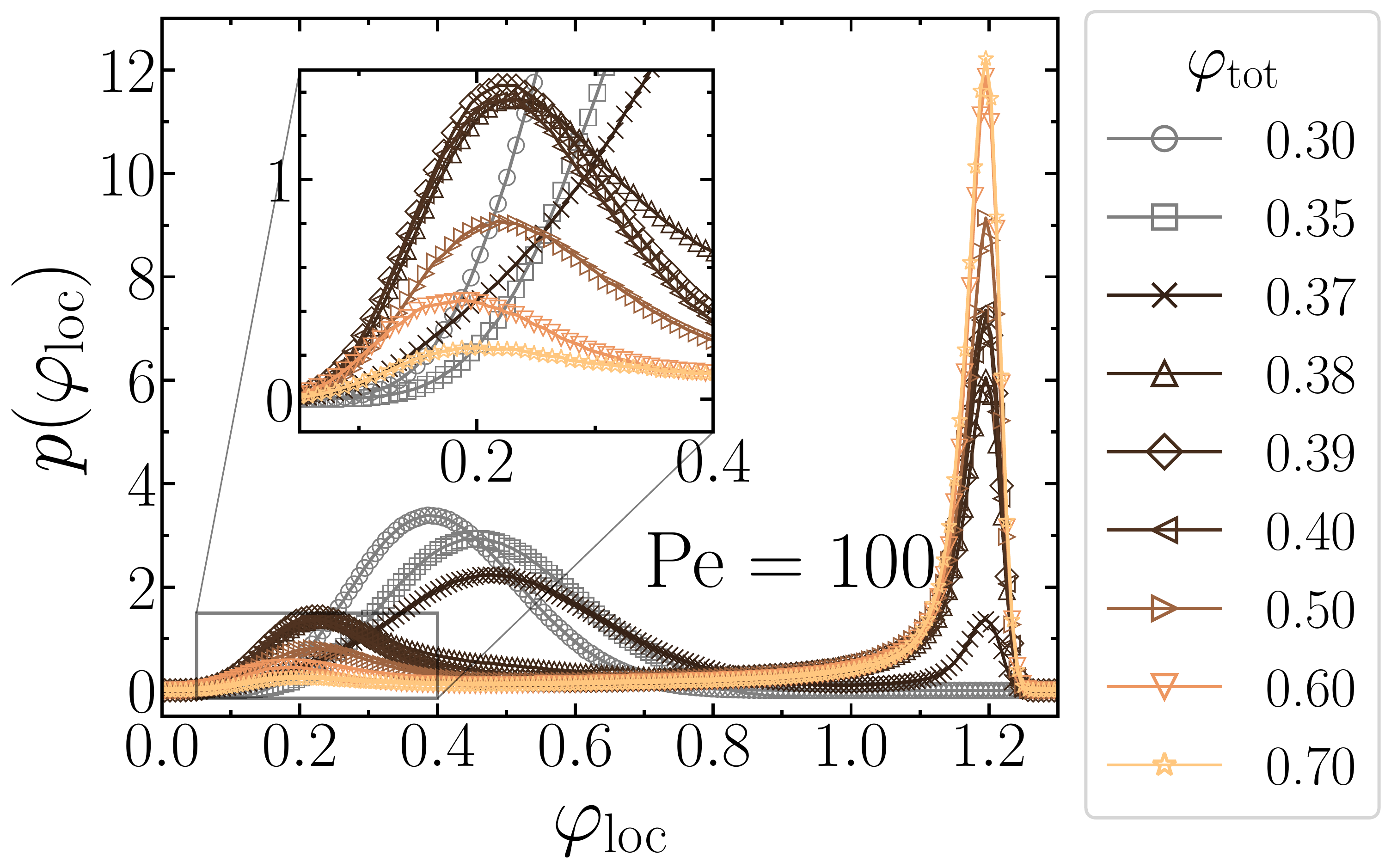}
	\caption{Distribution of the local area fraction for different $\varphi_{\rm tot}$ (values are shown in the key). Gray curves correspond to a uniform state, colored curves to a phase-separated (MIPS) state. Parameters: $N=20000$, $m/(\gamma_{\text{t}}\tau_{\rm p})=5\times 10^{-2}$,  $I/(\gamma_{\text{r}}\tau_{\rm p})=5\times10^{-3}$, $\epsilon/(k_{\text{B}}T_{\text{b}})=10$, $\sigma/\sqrt{D_{\rm r}D_{\rm t}}=1$, ${\rm Pe}=100$.}
	\label{fig:phi-loc-dist}
\end{figure}

To determine the phase diagram, we used a quadratic simulation area with periodic boundary conditions and $N=20000$ particles. We scanned the parameter ranges $\varphi_{\rm tot}\in [0.1,0.9]$ and ${\rm Pe}\in [0,300]$ and averaged over 3--10 realizations for each parameter combination resulting in about 4770 simulations in total. The phase transition line between the uniform and the coexistence (MIPS) regime was obtained based on the distribution of the local area fraction $p(\varphi_{\rm loc})$, which is unimodal in the uniform regime and bimodal in the coexistence regime. We calculated $p(\varphi_{\rm loc})$ based on averages over circles of radius $5\sigma$ and 3--10 realizations using the \verb|freud| Python library \cite{Ramasubramani_ComPhysComm_2020}. The results are exemplarily shown in Fig.\ \ref{fig:phi-loc-dist} for Pe=100. 

\section*{Density-Dependent Swimming Speed}
\begin{figure}
	\centering
	\includegraphics[width=0.6\linewidth]{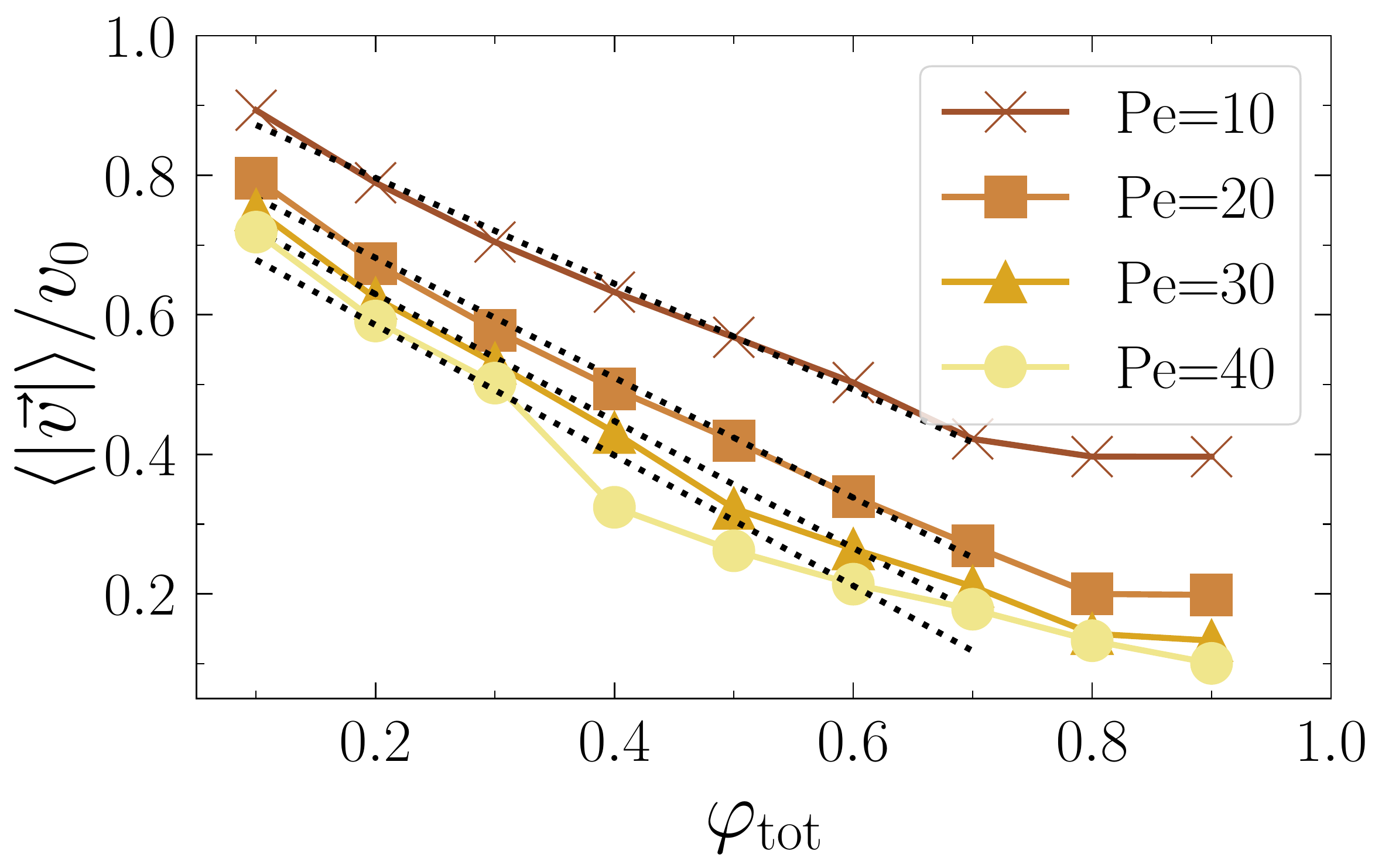}
	\caption{Density dependence of the mean speed (parameters as in Fig.\ \ref{fig:phi-loc-dist}) for different Pe (values are shown in the key). Black dotted lines are fits of $\langle |\vec{v}| \rangle/v_0=a(1-\varphi_{\rm tot}/\varphi^*)$ to the first seven data points. For Pe=10 and Pe=20, we get $a=0.95\pm0.01$, $\varphi^*=1.25\pm0.03$ and $a=0.85\pm0.02$, $\varphi^*=0.99\pm0.03$, respectively.}
	\label{fig:v-of-phi}
\end{figure}

To support our discussion about the counteracting and supportive feedback loop, we exemplarily evaluated the dependence of the mean speed $\langle |\vec{v}| \rangle$ on the total area fraction $\varphi_{\rm tot}$ for ${\rm Pe}\in\lbrace 10,20,30,40\rbrace$ (Fig.\ \ref{fig:v-of-phi}). When we have no MIPS (small Pe) at low enough area fractions, a linear dependence similar to the overdamped regime \cite{Cates_AnnuRevCondensMatterPhys_2015,Stenhammar_PhysRevLett_2013,Stenhammar_SoftMatter_2014} is observed, which breaks down at large area fractions. For higher Pe and especially in the MIPS regime, the linear dependence also breaks down as already found for overdamped ABPs \cite{Stenhammar_PhysRevLett_2013}, but $\langle |\vec{v}| \rangle$ is still decreasing with increasing $\varphi_{\rm tot}$. 

\begin{figure*}
	\centering
	\includegraphics[width=0.75\linewidth]{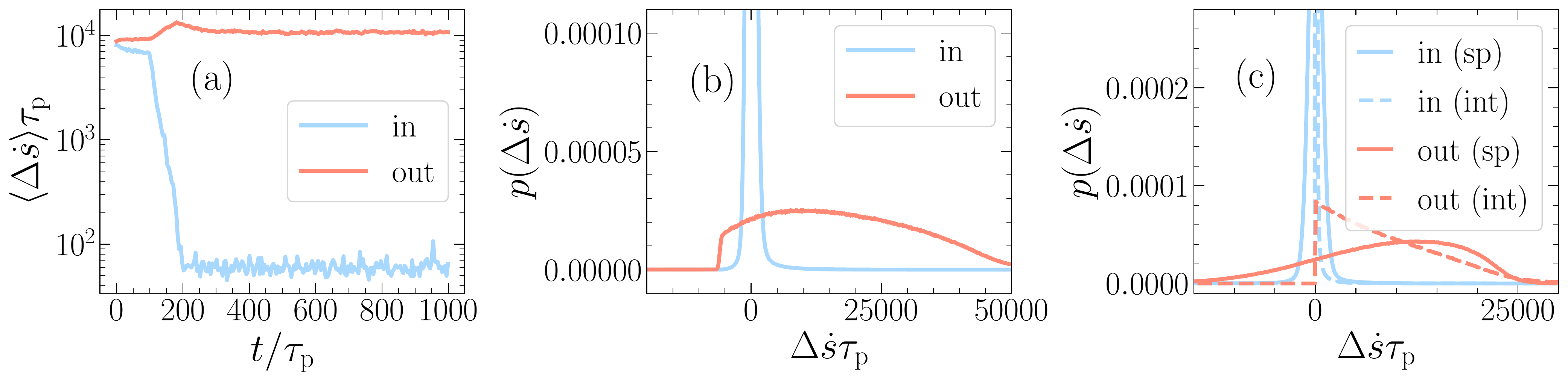}
	\caption{(a) Mean entropy production rate inside and outside the refrigerator domain over time. Time-averaged probability density of the entropy production rate (b) and of the entropy production rate separated in the self-propulsion (sp) contribution and the interaction (int) contribution (c). Parameters: $N=64000$, $\varphi_{\rm tot}=0.35$, Pe$_{\rm in}=105$, Pe$_{\rm out}=110$, $m/(\gamma_{\text{t}}\tau_{\rm p})=5\times 10^{-2}$, $I/(\gamma_{\text{r}}\tau_{\rm p})=5\times10^{-3}$, $\epsilon/(k_{\text{B}}T_{\text{b}})=10$, $\sigma/\sqrt{D_{\rm r}D_{\rm t}}=1$, $x_0/L_x=0.1$.}
	\label{fig:figs3}
\end{figure*}

\section*{Concept of effective temperature and heat transfer}
Active systems, which consist of self-propelled particles, are intrinsically out of equilibrium. Hence, the second law of thermodynamics only applies to the overall system (particle plus fluid/substrate) but not to the particle subsystem alone. Therefore, from a microscopic viewpoint, cooling down the active particles locally without transferring heat to an external (spatially separated) bath means that heat is transported from the active particles to the surrounding solvent. The latter has a comparatively large number of degrees of freedom and would heat up only very little (or very slowly) while the active particles cool down by orders of magnitude. In this work, we define the temperature of the active particles in terms of their (translational) kinetic energy. In equilibrium, this kinetic temperature is equal to the thermodynamic temperature as long as the Hamiltonian of the system is quadratic in the momentum coordinates (equipartition theorem) \cite{Schroeder_Book_AnIntroductionToThermalPhysics_2000}. It can be shown that the kinetic temperature is also equivalent to the virial temperature even for active Brownian particles \cite{Mandal_PhysRevLett_2019}. Note that a temperature based on fluctuation-dissipation relations (FDRs) can only be defined by generalizing the equilibrium FDR, which is violated in active systems \cite{Petrelli_PhysRevE_2020,Cugliandolo_FluctNoiseLett_2019,Caprini_Symmetry_2021,Levis_EPL_2015,Szamel_PhysRevE_2014,Loi_PhysRevE_2008,Puglisi_PhysRep_2017}. However, one can construct a 'thermometer' that measures an effective temperature of the active-particle subsystem as discussed below.

\section*{Entropy Production}
The entropy production rate measures how strongly detailed balance is broken and thus, how far the state of a system deviates from an equilibrium state \cite{Fodor_PhysRevLett_2016,Chaudhuri_PhysRevE_2014,Cengio_JStatMech_2021,Seifert_RepProgPhys_2012,OByrne_NatRevPhys_2022}. Therefore, entropy production is required to observe a temperature difference between coexisting phases in a steady state (however, the opposite is not true \cite{Ro_ArXiv_2022,Mandal_PhysRevLett_2019}). The entropy production rate for inertial active Brownian particles can be calculated as follows: let $\Gamma$ denote one trajectory of the system, i.e., the set of positions, velocities, and orientation angles for all particles over a time interval $[0,t]$. Furthermore, we denote the corresponding time-reversed trajectory by $\tilde{\Gamma}$. The entropy production is defined by \cite{Shankar_PhysRevE_2018}
\begin{equation}
	\Delta s=\ln\left[\frac{p(\Gamma)}{p(\tilde{\Gamma})}\right],\label{eq:entropy-production}
\end{equation}
where $p(\Gamma)$ denotes the probability density of the trajectory $\Gamma$, which is given by the Onsager-Machlup functional \cite{Onsager_PhysRev_1953}. For underdamped ABPs, we obtain
\begin{equation}
	p(\Gamma)\propto\exp\left\lbrace-\frac{\gamma_{\rm t}}{4k_{\rm B}T_{\rm b}}\sum_{i=1}^{N}\int_{0}^{t}{\rm d}\tau\left[\frac{m}{\gamma_{\rm t}}\dot{\vec{v}}_i+\vec{v}_i-v_0\hat{p}_i-\frac{1}{\gamma_{\rm t}}\vec{F}_{{\rm int},i}\right]^2\right\rbrace,\label{eq:onsager-machlup-result}
\end{equation}
where $\vec{F}_{{\rm int},i}=\sum_{j=1,j\neq i}^{N}\nabla_{\vec{r}_i}u\left(r_{ij}\right)$ denotes the interaction force due to the WCA potential with $r_{ij}=|\vec{r}_i-\vec{r}_j|$ [see Eq.\ (\ref{eq:wca})]. Following Refs.\ \cite{Szamel_PhysRevE_2019,Shankar_PhysRevE_2018,Pietzonka_JPhysAMathTheor_2018,Nemoto_PhysRevE_2019,Cagnetta_PhysRevLett_2017,GrandPre_PhysRevE_2021}, positions, velocities, orientations, and forces transform under time reversal as $\vec{r}(t)=\vec{r}(-t)$, $\vec{v}(t)=-\vec{v}(-t)$, $\hat{p}(t)=\hat{p}(-t)$, and $\vec{F}_{\rm int}(t)=\vec{F}_{\rm int}(-t)$, respectively. Therefore, we finally obtain the total entropy production
\begin{equation}
	\Delta s=\frac{1}{D_{\rm t}}\sum_{i=1}^{N}\int_{0}^{t}{\rm d}\tau\,\left[v_0\hat{p}_i\cdot\vec{v}_i+\frac{1}{\gamma_{\rm t}}\vec{v}_i\cdot\vec{F}_{{\rm int},i}-\frac{m}{\gamma_{\rm t}}\vec{v}_i\cdot\dot{\vec{v}}_i\right],\label{eq:entropy-production-final}
\end{equation}
with $D_{\rm t}=k_{\rm B}T_{\rm b}/\gamma_{\rm t}$. Since the last term obeys $2\vec{v}_i\cdot\dot{\vec{v}}_i=\partial_t\vec{v}_i^{~2}$ and the mean kinetic energy $\langle m\vec{v}^{~2}/2\rangle$ is constant in the steady state, this term vanishes. Thus, we finally have two contributions: one from the self propulsion and one from pair interactions. Hence, the mean entropy production rate is given by
\begin{equation}
	\langle\Delta\dot{s}\rangle=\frac{1}{ND_{\rm t}}\sum_{i=1}^{N}\left[v_0\hat{p}_i\cdot\vec{v}_i+\frac{1}{\gamma_{\rm t}}\vec{v}_i\cdot\vec{F}_{{\rm int},i}\right].\label{eq:entropy-production-rate}
\end{equation}
Its time evolution is shown in Fig.\ \ref{fig:figs3}(a): once the steady state is reached, the entropy production rate in the refrigerator domain is about two orders of magnitude smaller than in the environment. The distribution of the entropy production rate is narrow and centered around a small positive value for particles inside the refrigerator and broad for particles in the environment [Fig.\ \ref{fig:figs3}(b)]. The two contributions are demonstrated in Fig.\ \ref{fig:figs3}(c) confirming our observations.

\begin{figure}
	\centering
	\includegraphics[width=0.9\linewidth]{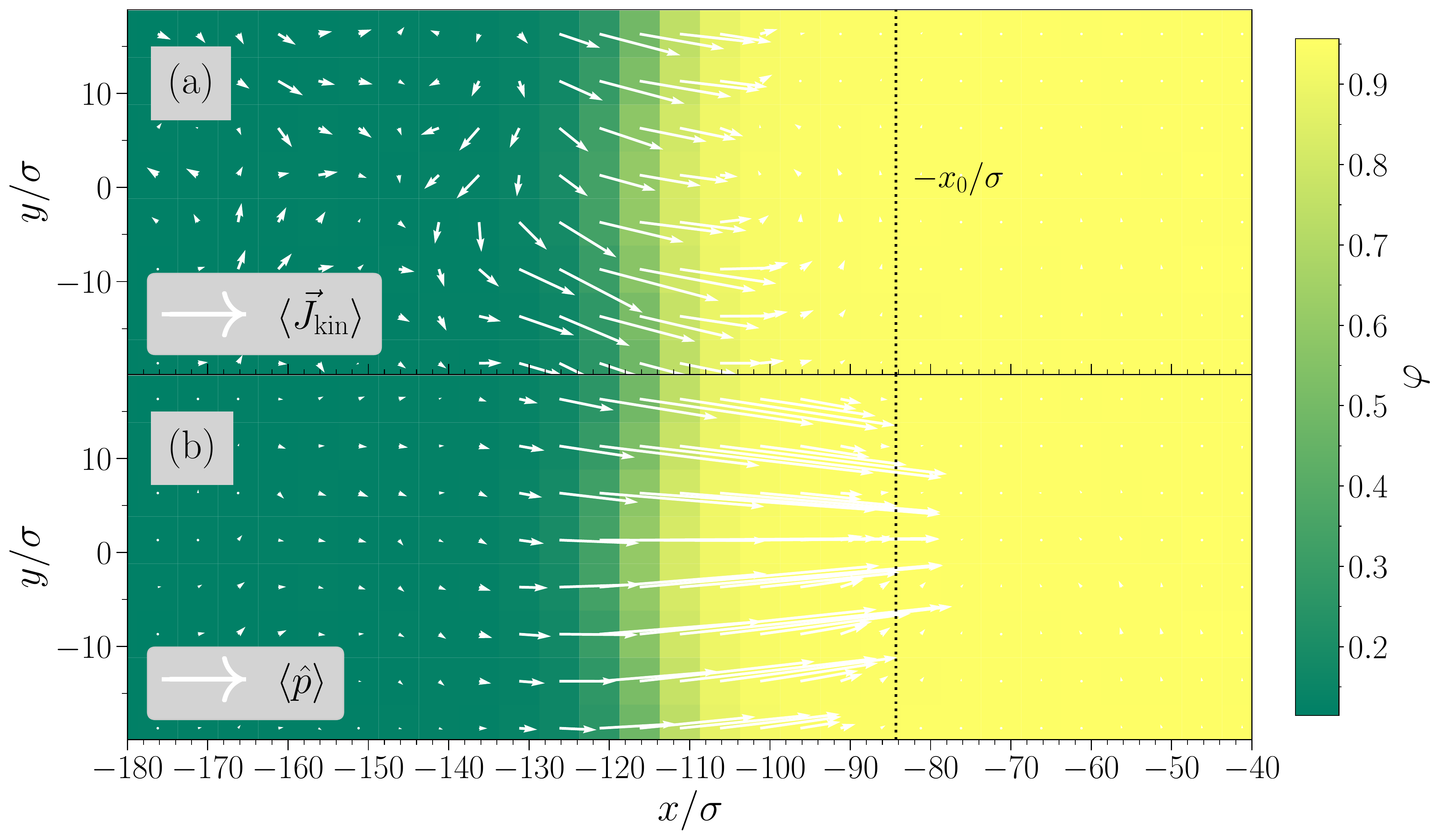}
	\caption{(a) Flow of kinetic energy [see Eq.\ (\ref{eq:ekinflow-particle})]. The white arrows denote the direction of the flow, their length correspond to the strength in arbitrary units. (b) Polarization field. The white arrows denote the direction, their length denotes the strength of the polarization in arbitrary units. The background colors show the area fraction. All data are averaged over time in the steady state and over 20 realizations. Parameters: $N=16000$, $\varphi_{\rm tot}=0.35$, Pe$_{\rm in}=105$, Pe$_{\rm out}=110$, $m/(\gamma_{\text{t}}\tau_{\rm p})=5\times 10^{-2}$, $I/(\gamma_{\text{r}}\tau_{\rm p})=5\times10^{-3}$, $\epsilon/(k_{\text{B}}T_{\text{b}})=10$, $\sigma/\sqrt{D_{\rm r}D_{\rm t}}=1$, $x_0/L_x=0.1$.}
	\label{fig:figs4}
\end{figure}

\section*{Kinetic Temperature Gradient and Energy Flow of the Active Particles}
Let us now first discuss the flow of kinetic energy at the level of the active particles and then the energy flow within the bath (liquid/gas) that surrounds the particles. As we will see, completely consistent with the basic thermodynamic fact, energy naturally flows from hot to cold regions both at the level of the active particles and within the bath. This energy flow persists in steady state and is maintained by the driving (e.g., due to a laser).

We calculated the kinetic energy flow over the boundary of the dense phase, which can be defined by
\begin{equation}
	\vec{J}_{{\rm kin}}(\vec{r})=\frac{1}{2}m\langle\vec{v}(\vec{r})\rangle^{~2}\rho_{\rm loc}(\vec{r})\langle \vec{v}(\vec{r})\rangle, \label{eq:ekinflow-particle}
\end{equation}
where $\rho_{\rm loc}(\vec{r})$ denotes the (local) particle number density and $\vec{v}(\vec{r})$ the velocity of the ABPs (averaged over a small area of size $\Delta x\Delta y$ with $\Delta x=\Delta y=5\sigma$). The result is demonstrated in Fig.\ \ref{fig:figs4}(a): as expected, an inward flow of kinetic energy is observed at the boundary of the dense phase, which is mainly caused by a local alignment of the effective self-propulsion force as demonstrated by the coarse-grained polarization field $\langle\hat{p}\rangle$ shown in Fig.\ \ref{fig:figs4}(b). This kind of alignment has already been observed for overdamped ABPs in a motility gradient \cite{Soeker_PhysRevLett_2021,Auschra_PhysRevE_2021}.

\begin{figure}
	\centering
	\includegraphics[width=0.7\linewidth]{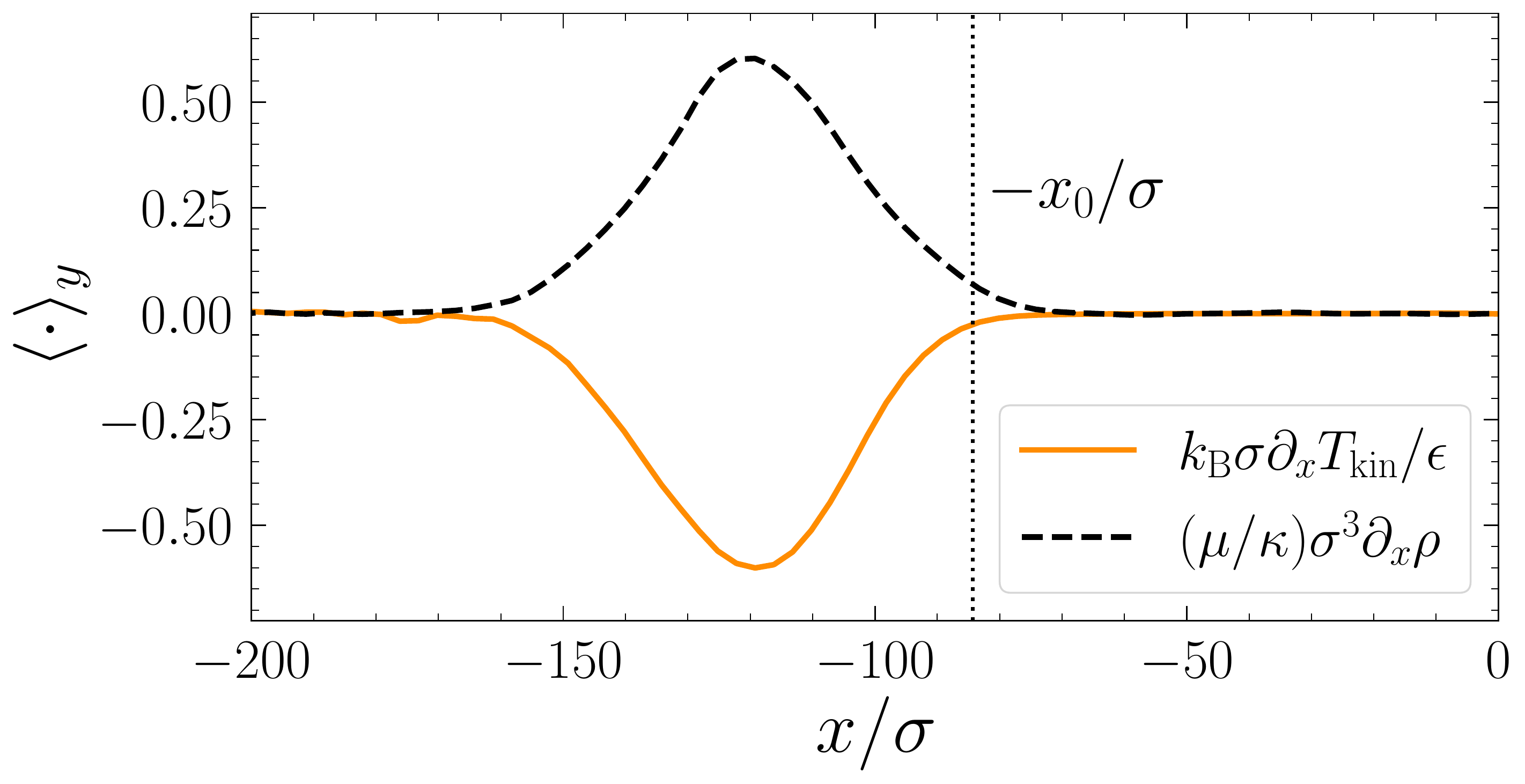}
	\caption{Gradient of the kinetic temperature $T_{\rm kin}$ and the particle density $\rho$ at the border of the cooling domain in the steady state (averaged over the $y$ direction and over 20 realizations). The gradient of the particle density is scaled with a factor $\mu/\kappa\approx 23$ (parameters as in Fig.\ \ref{fig:figs4}).}
	\label{fig:figs5}
\end{figure}

Furthermore, Fourier's law can be used to relate the kinetic energy flow to a (kinetic) temperature gradient:
\begin{equation}
	\vec{J}_{\rm Fourier}=-\kappa\nabla T_{\rm kin},
\end{equation}
where $\kappa$ denotes an effective thermal conductivity \cite{Zwanzig_Book_NonequilibriumStatisticalMechanics_2001}. The temperature gradient must be compensated by a particle flux in the steady state in the presence of a density gradient \cite{Komatsu_PhysRevX_2015,Brey_PhysRevE_1998} and the condition
\begin{equation}
	-\kappa\nabla T_{\rm kin}-\mu\nabla\rho=0\label{eq:steadystatecondition}
\end{equation}
with a positive transport coefficient $\mu$ and particle density $\rho$ must hold. In particular, the (kinetic) temperature and density gradients are opposite at the border of the refrigerator domain and Eq.\ (\ref{eq:steadystatecondition}) is fulfilled in our simulations once we set $\mu/\kappa\approx 23$ (see Fig.\ \ref{fig:figs5}).

\section*{Bath Temperature and Bath Energy Flow}
\begin{figure}
	\centering
	\includegraphics[width=0.9\linewidth]{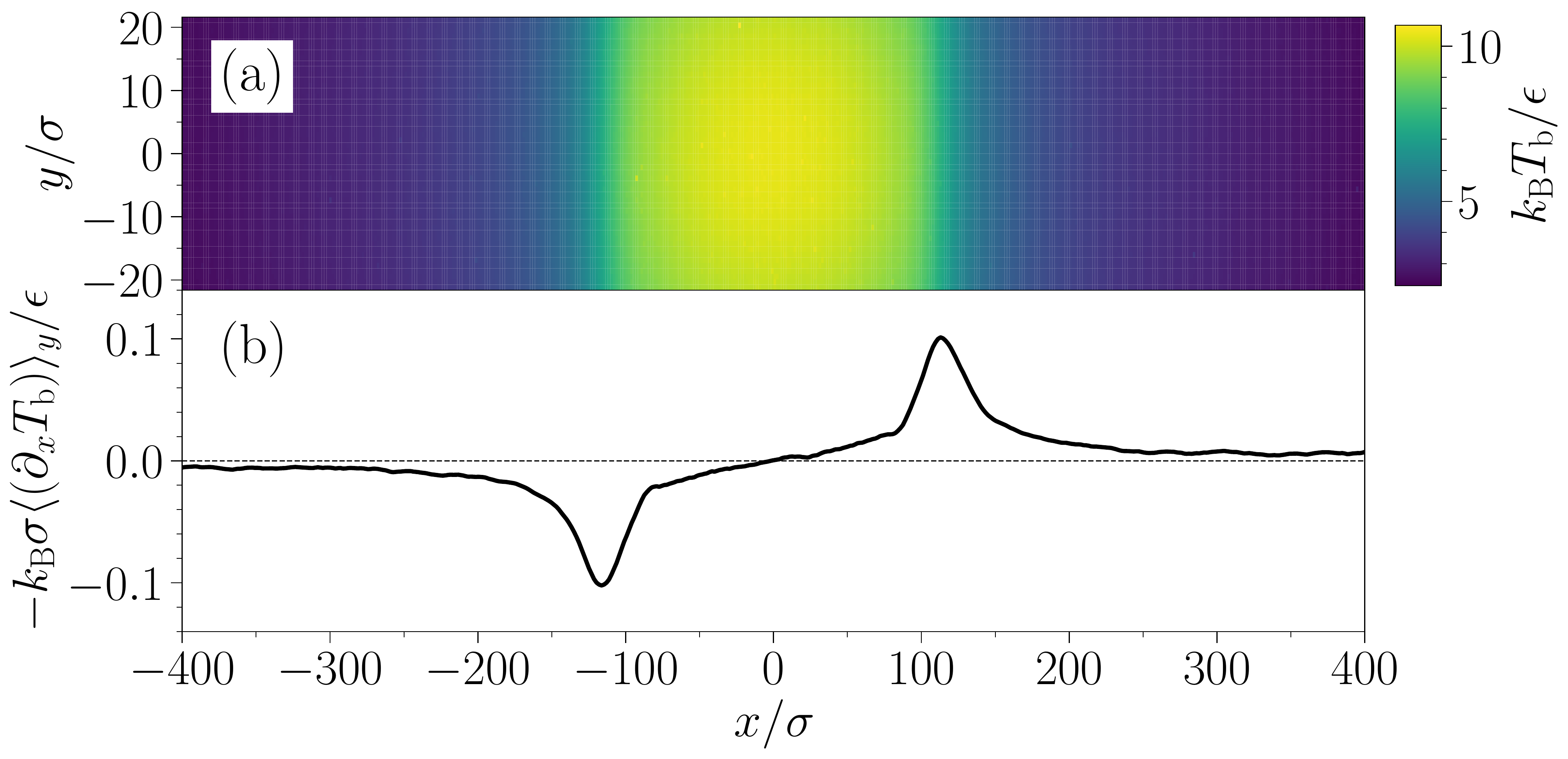}
	\caption{Estimated steady-state bath temperature field (a) and its negative gradient in $x$ direction averaged over the $y$ coordinate (b) for a small Pe difference ${\rm Pe}_{\rm out}-{\rm Pe}_{\rm in}=5$. The color denotes the reduced temperature from dark blue (cold) to yellow (hot). Parameters: Pe$_{\rm in}$=105, Pe$_{\rm out}$=110, $\alpha\tau_{\rm p}/\sigma^2=1.0$, $g_0/(T_{\rm b}\sigma^2)=10^{-4}$.}
	\label{fig:temp-field}
\end{figure}

\begin{figure}
	\centering
	\includegraphics[width=0.9\linewidth]{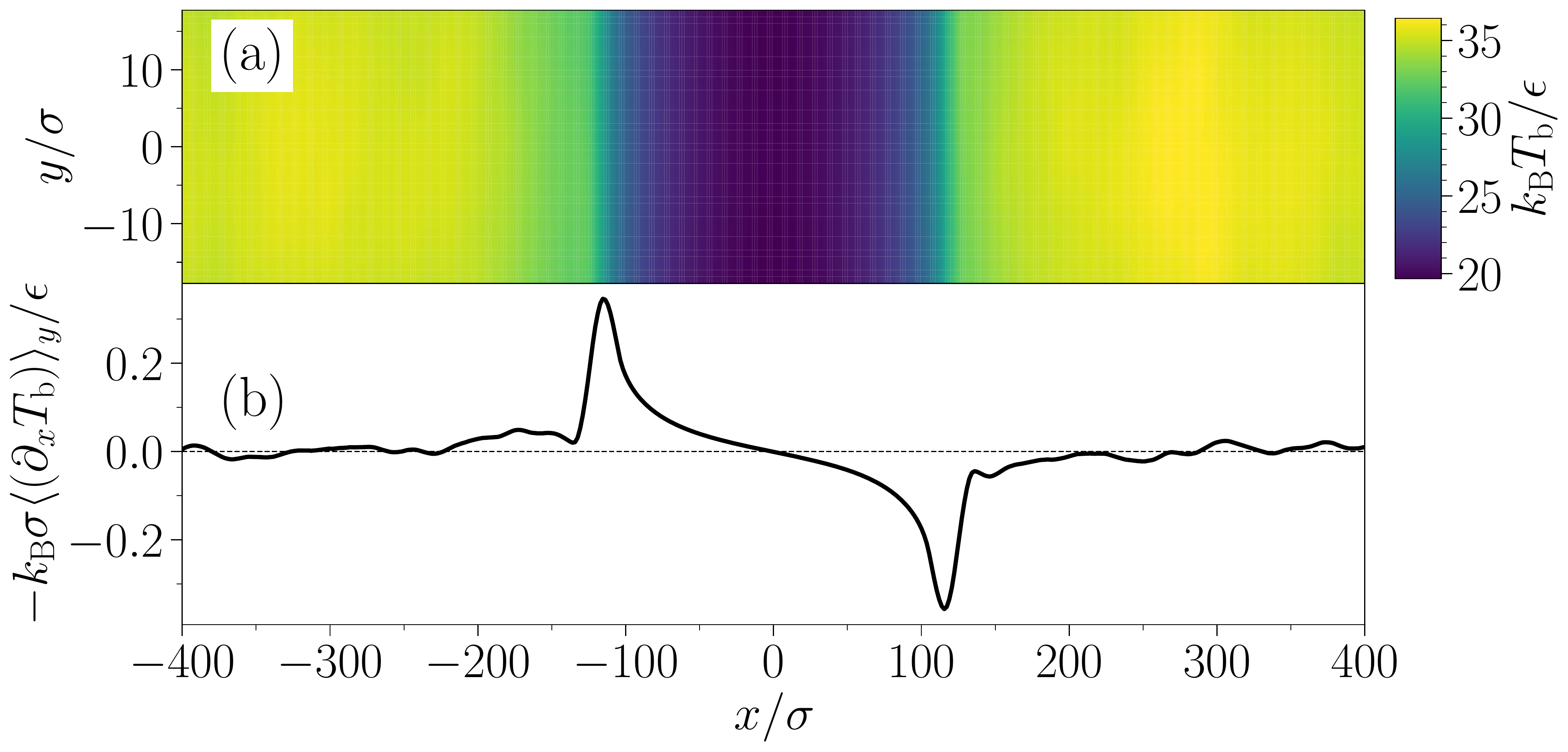}
	\caption{Estimated steady-state bath temperature field (a) and its negative gradient in $x$ direction averaged over the $y$ coordinate (b) for a large Pe difference ${\rm Pe}_{\rm out}-{\rm Pe}_{\rm in}=360$. The color denotes the reduced temperature from dark blue (cold) to yellow (hot). Parameters: Pe$_{\rm in}$=40, Pe$_{\rm out}$=400, $\alpha\tau_{\rm p}/\sigma^2=1.0$, $g_0/(T_{\rm b}\sigma^2)=10^{-4}$.}
	\label{fig:temp-field-reversed}
\end{figure}

We now complement the discussion regarding the bath temperature in the main text with a minimal model, which explicitly shows that the energy persistently flows from hot to cold within the bath in the steady state. We begin with the heat equation
\begin{equation}
	\frac{\partial T_{\rm b}}{\partial t}=\alpha\nabla^2T_{\rm b}+g(\vec{r},t)\label{eq:heat-equation}
\end{equation}
with bath temperature field $T_{\rm b}(\vec{r},t)$, thermal diffusivity $\alpha$, and heat source or sink $g(\vec{r},t)$ \cite{Livi_Book_NonequilibriumStatisticalPhysics_2017,Bird_Book_TransportPhenomena_2002,Lebon_Book_UnderstandingNonEquilibriumThermodynamics_2008,Falasco_PhysRevE_2014}. Here, for simplicity, we assume that heat diffusion dominates over heat advection and neglect the latter. Describing each ABP as a point-like heat source in 3D (which is confined to a 2D interface/substrate) for simplicity with strength proportional to its self-propulsion speed $v_0$ (reasonable for laser-powered Janus particles for example \cite{Jiang_PhysRevLett_2010}), we can write $g(\vec{r})=g_0\sum_{i=1}^{N}v_{0,i}\delta(\vec{r}-\vec{r}_i)$ with self-propulsion speed $v_{0,i}$ of the $i$-th particle and a suitable constant $g_0$. Here, we assume that all the energy which is absorbed by an active particle from the (external) energy source is ultimately transferred to the bath if we neglect temperature changes of the particle material. The corresponding solution of Eq.\ (\ref{eq:heat-equation}) in the steady state ($\partial_t T_{\rm b}=0$) can be written in terms of the Greens function \cite{Jentschura_JPhysComm_2018,Liebchen_ArXiv_2019} as 
\begin{equation}
	T_{\rm b}(\vec{r})=\frac{g_0}{4\pi\alpha}\sum_{i=1}^{N}\frac{v_{0,i}}{|\vec{r}-\vec{r}_i|}.\label{eq:bath-temp-model-sim}
\end{equation}
Based on this minimal model, we estimate the steady-state temperature field of the bath by inserting the coordinates of the active particles into Eq.\ (\ref{eq:bath-temp-model-sim}) and averaging over 20 snapshots in the steady state (Figs.\ \ref{fig:temp-field} and \ref{fig:temp-field-reversed}). For a uniform Pe or a small Pe difference, it turns out that regions of high ABP density feature a higher bath temperature (yellow) than regions of low ABP density [blue, see Fig.\ \ref{fig:temp-field}(a)]. Consequently, we observe an energy flow from the dense region to the dilute region within the bath [Fig.\ \ref{fig:temp-field}(b)], which is related to the temperature field by Fourier's law \cite{Zwanzig_Book_NonequilibriumStatisticalMechanics_2001,Bird_Book_TransportPhenomena_2002,Livi_Book_NonequilibriumStatisticalPhysics_2017,Lebon_Book_UnderstandingNonEquilibriumThermodynamics_2008,Cole_Book_HeatConductionUsingGreensFunctions_2011}. That is, in Fig.\ \ref{fig:temp-field}(b), we have an energy current pointing to the left for $x<0$ (where -$\partial_x T_{\rm b}<0$) and an energy current pointing to the right for $x>0$ (where $-\partial_x T_{\rm b}>0$). This energy current can be reversed by choosing a large Pe difference (${\rm Pe}_\text{out} \gg {\rm Pe}_\text{in}$) as shown in Fig.\ \ref{fig:temp-field-reversed}(a) and (b).

While we have discussed the minimal model for a finite number of particles so far, in the thermodynamic limit, one needs to take into account that heat would be absorbed by boundaries or would ultimately be radiated off the system, which we need to take into account to obtain a converged temperature field. A minimal way to achieve convergence is to introduce a loss term $-k_{\rm d} T_{\rm b}$ with some suitable loss coefficient $k_{\rm d}$. For convenience, we also introduce the spatially dependent self-propulsion speed $v_0(\vec{r})$ and the steady-state particle density $\rho(\vec{r})$. Then, the heat source reads $g(\vec{r})=g_0v_0(\vec{r})\rho(\vec{r})$ and the steady-state heat equation reads
\begin{equation}
	0=\alpha\nabla^2T_{\rm b}+g(\vec{r})-k_{\rm d}T_{\rm b}.
\end{equation}
It's solution can again be written in terms of a Greens function \cite{Economou_Book_GreensFunctionsInQuantumPhyiscs_2006} as
\begin{equation}
	T_{\rm b}(\vec{r})=\frac{g_0}{4\pi\alpha}\int{\rm d}^3r\,'\,v_0(\vec{r}\,')\rho(\vec{r}\,')\frac{{\rm e}^{-\sqrt{\frac{k_{\rm d}}{\alpha}}|\vec{r}-\vec{r}\,'|}}{|\vec{r}-\vec{r}\,'|}.
\end{equation}
This shows that the bath temperature is high in regions where the product $v_0(\vec{r}\,') \rho(\vec{r}\,')$ is large. That is, for a small Pe difference, $T_{\rm b}$ is large in regions of high particle density and hence, according to Fourier's law, heat is flowing away from such regions within the bath [Fig.\ \ref{fig:temp-field}]. In contrast, heat will flow from the dilute to the dense region within the bath for ${\rm Pe}_{\rm out}\gg {\rm Pe}_{\rm in}$ [Fig.\ \ref{fig:temp-field-reversed}].


\section*{Thermometer for Active Particles}
\begin{figure}[h]
	\centering
	\includegraphics[width=0.7\linewidth]{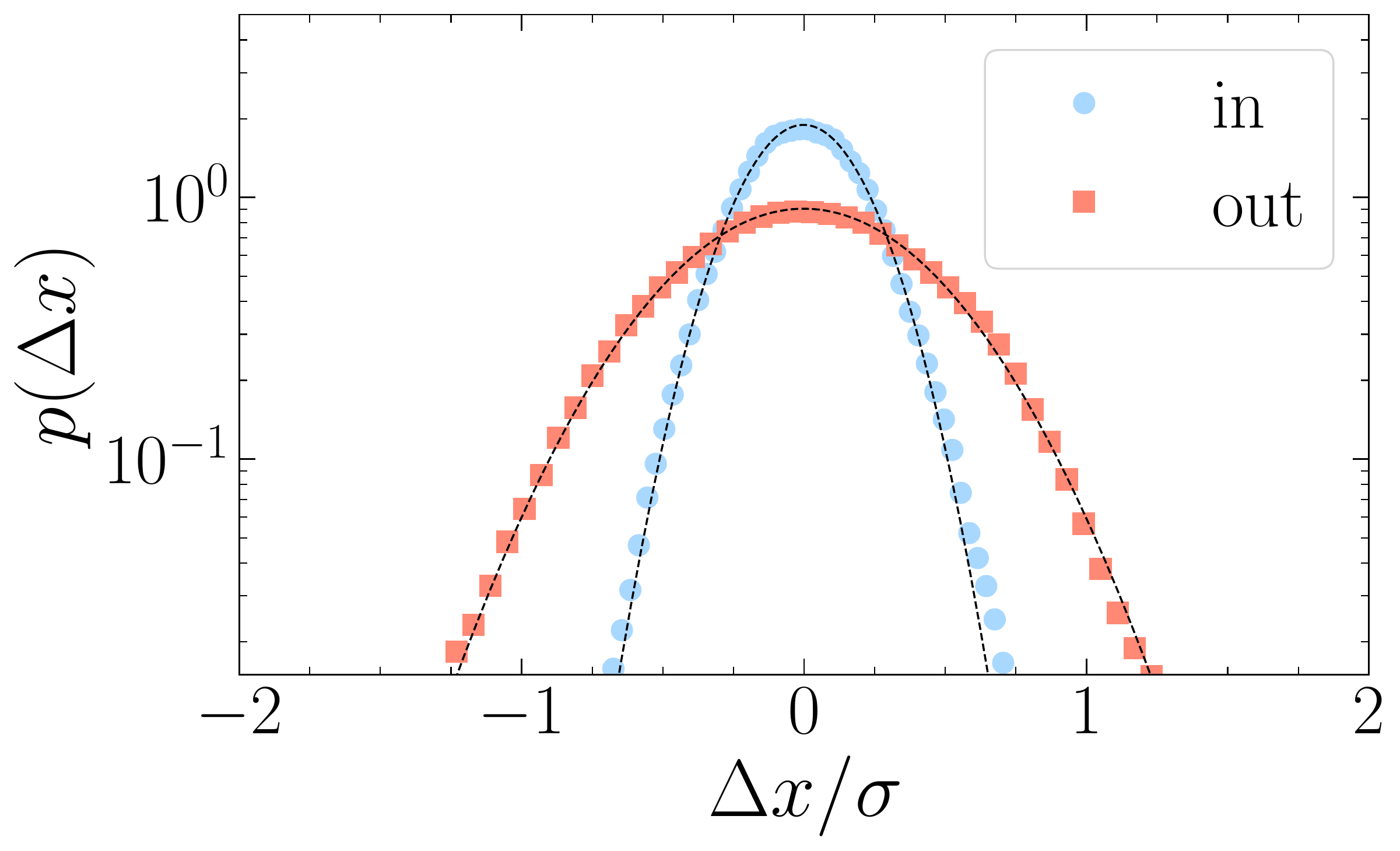}
	\caption{Displacement distribution of passive tracers with mass $M/(\gamma_{\rm t}\tau_{\rm p})=1.0$ and radius $R/\sigma=0.005$ trapped in a harmonic potential of strength $k\sigma^2/\epsilon=60$ inside and outside the cooling domain averaged over 40 realizations. Black dashed lines are Gaussian fits yielding  $k_{\rm B}T_{\rm AP}^{\rm (in)}/\epsilon\approx 2.65$ and $k_{\rm B}T_{\rm AP}^{\rm (out)}/\epsilon\approx 11.03$. Parameters: $N=16000$, $\varphi_{\rm tot}=0.35$, Pe$_{\rm in}=105$, Pe$_{\rm out}=110$, $m/(\gamma_{\text{t}}\tau_{\rm p})=5\times 10^{-2}$, $I/(\gamma_{\text{r}}\tau_{\rm p})=5\times10^{-3}$, $\epsilon/(k_{\text{B}}T_{\text{b}})=10$, $\sigma/\sqrt{D_{\rm r}D_{\rm t}}=1$, $x_0/L_x=0.1$.}
	\label{fig:thermometer-dx-dist}
\end{figure}

A standard thermometer would measure the temperature of the surrounding bath. Here, we propose a 'thermometer' assigning a temperature to the active particles based on passive tracer particles trapped in a harmonic potential \text{$U_{\rm harm.}(\vec{r})=k\vec{r}^{~2}/2$} of strength $k$. As we will see, the temperature which this thermometer measures behaves analogously to the kinetic temperature, which we discuss in the main text. The tracer particles could be made semi-permeable experimentally as in Refs.\ \cite{Staedler_Nanoscale_2009,Chang_Science_1964}, so that they essentially interact only with the active particles. The distribution of tracer displacements $\Delta x$ along the $x$-axis is expected to be Gaussian and is found to be Gaussian in our simulations (similar results are obtained for the displacements $\Delta y$ along the $y$ axis). Its variance $\langle(\Delta x-\langle\Delta x\rangle)^2 \rangle$ is used to estimate an effective temperature
\begin{equation}
	k_{\rm B}T_{\rm AP}(k)=k\left\langle\left(\Delta x-\langle\Delta x\rangle\right)^2 \right\rangle,
\end{equation}
of the active particles, which generally depends on the strength $k$ of the harmonic potential \cite{Greinert_PhysRevLett_2006,Ye_SoftMatter_2020,Demery_JStatMech_2019,Maggi_PhysRevLett_2014}. Obtaining a consistent measurable value for $T_{\rm AP}$ is however not completely straight forward: first, the tracer particles should be small because in the dense phase, large tracers would be trapped by surrounding active particles. Second, the tracer particles should also be sufficiently heavy such that they do not slow down too much between subsequent collisions. Third, $k$ has to be large because especially a tracer particle in the dense phase should only move within the cage of the surrounding active particles (if $k$ is too large, however, collisions with the active particles are too rare on the time scale of the simulation). Accordingly, we use tracer particles with mass $M/(\gamma_{\rm t}\tau_{\rm p})=1.0$ and radius $R/\sigma=0.005$ and place one tracer in the middle of the cooling domain and one in the dilute phase. For an exemplary value $k\sigma^2/\epsilon=60$ we obtain a low effective temperature $k_{\rm B}T_{\rm AP}^{\rm (in)}/\epsilon\approx 2.65$ inside the cooling domain and a high temperature $k_{\rm B}T_{\rm AP}^{\rm (out)}/\epsilon\approx 11.03$ outside the cooling domain (see Fig.\ \ref{fig:thermometer-dx-dist}). Importantly, a lower temperature is measured inside the cooling domain for all values of $k$. This is consistent with our findings based on the kinetic temperature.

\section*{Role of the Refrigerator Size}
The length $x_0$ defines the size of the targeted cooling domain (cf.\ Fig.\ 1(a) of the main text). The role of $x_0$ for our proposed cooling mechanism and the regimes (I) and (II) (see Fig.\ 1(b) of the main text) can be understood as follows:

\vspace{0.2cm}
\emph{Regime (I):} due to the counteracting feedback loop, the density inside the refrigerator region decreases below $\varphi_{\rm tot}$ and prevents the particles from undergoing motility-induced phase separation (MIPS). Simultaneously, the density outside the refrigerator region increases. The steady-state density in the environment of the refrigerator region strongly depends on the value of $x_0$: for $x_0$ comparable to the system size (i.e., $L_x$), the number of particles which can leave the refrigerator region due to the counteracting feedback loop is large and vice versa. Thus, the steady-state density outside the active refrigerator increases with increasing $x_0$. Due to the linear dependence of the mean speed on the area fraction, the kinetic temperature outside the refrigerator decreases with increasing $x_0$ and causes a (weak) cooling of the environment.

\begin{figure}[h]
	\centering
	\includegraphics[width=0.8\linewidth]{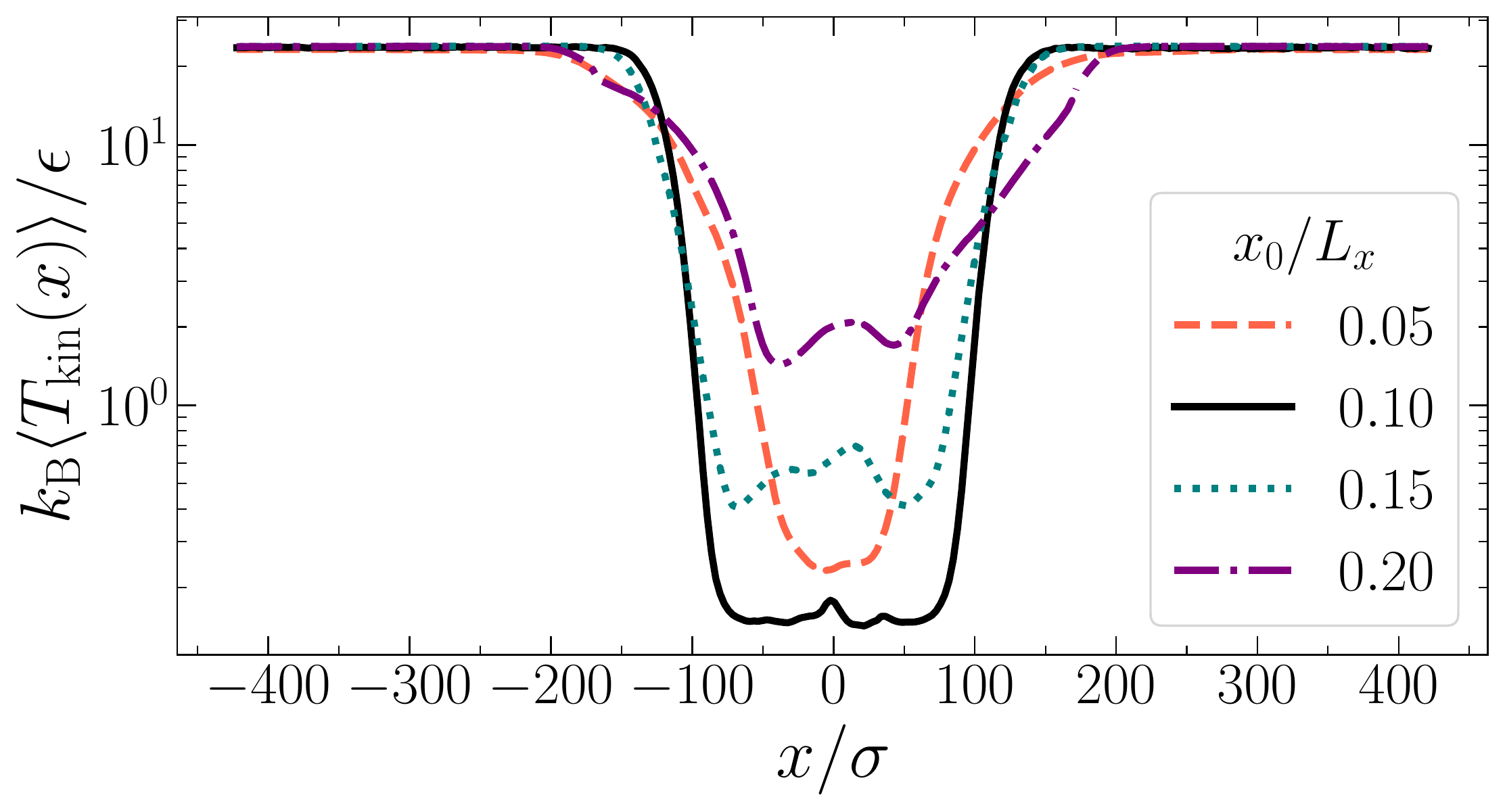}
	\caption{Time-averaged kinetic temperature profile for different sizes $x_0/L_x$ of the refrigerator domain (values are shown in the key) obtained from simulations with $N=16000$ particles, Pe$_{\rm in}=105$, Pe$_{\rm out}=110$, and $\varphi_{\rm tot}=0.35$ and averaged over 20 realizations. Further parameters: $m/(\gamma_{\text{t}}\tau_{\rm p})=5\times 10^{-2}$, $I/(\gamma_{\text{r}}\tau_{\rm p})=5\times10^{-3}$, $\epsilon/(k_{\text{B}}T_{\text{b}})=10$, $\sigma/\sqrt{D_{\rm r}D_{\rm t}}=1$.}
	\label{fig:figs6}
\end{figure}

\vspace{0.2cm}
\emph{Regime (II):} here, the cooling is triggered by MIPS inside the targeted cooling domain. For $x_0$ small compared to $L_x$, the dense cluster fills the whole target domain. Thus, particles are cooled in the whole cooling domain (black solid line in Fig.\ \ref{fig:figs6}). However, if $x_0/L_x\lesssim 0.05$, the dense cluster might occupy a spatial region larger than the refrigerator domain. Furthermore, the localization is less effective in this case such that the dense phase moves around the refrigerator domain (and eventually leaves it partially). As a consequence, the ensemble-averaged kinetic temperature inside the refrigerator is slightly larger as shown with the dashed red line in Fig.\ \ref{fig:figs6}. For $x_0/L_x\gtrsim0.15$, the dense cluster might not fill the whole domain anymore and it will be placed at a random position inside the domain, which finally decreases the cooling effect when taking the ensemble average (see, e.g., dash-dotted purple line in Fig.\ \ref{fig:figs6}). In the limit of very large $x_0$, i.e., $x_0/L_x\rightarrow 1$, the dense cluster is placed at a random position inside the cooling domain causing the ensemble-averaged temperature profile to be approximately uniform (Fig.\ 2(a) of the main text).

\section*{Robustness Against the System Size}
\begin{figure}[h]
	\centering
	\includegraphics[width=0.7\linewidth]{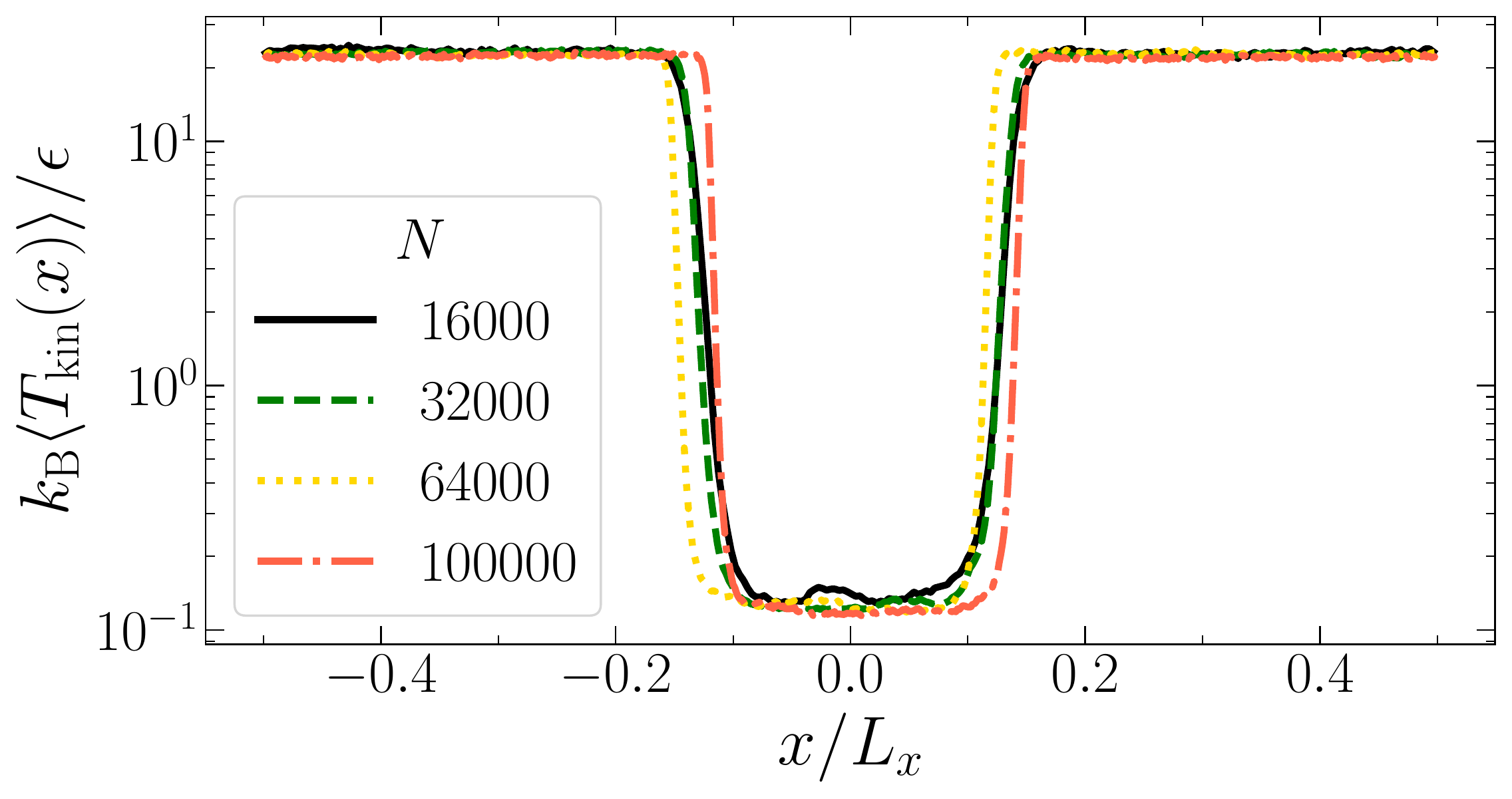}
	\caption{Time-averaged kinetic temperature profile in the steady state for the active refrigerator setup (see regime (II) in Fig.\ 1(b) of the main text) and different system sizes. The ratios $L_y/L_x=0.05$ and $x_0/L_x=0.1$ are kept constant as well as the total area fraction $\varphi_{\rm tot}=0.35$ while we varied the number of particles $N$ (values are shown in the key). Further parameters: Pe$_{\rm in}=105$, Pe$_{\rm out}=110$, $m/(\gamma_{\text{t}}\tau_{\rm p})=5\times 10^{-2}$, $I/(\gamma_{\text{r}}\tau_{\rm p})=5\times10^{-3}$, $\epsilon/(k_{\text{B}}T_{\text{b}})=10$, $\sigma/\sqrt{D_{\rm r}D_{\rm t}}=1$.}
	\label{fig:figs7}
\end{figure}

To ensure that our results are not affected by finite-size effects, we performed additional simulations with $N=32,000$, 64,000, and 100,000 particles by keeping the total area fraction and the ratio $x_0/L_x$ constant. As we show in Fig.\ \ref{fig:figs7}, we get essentially the same results for all studied system sizes resulting in a well defined refrigerator domain with a temperature difference of about two orders of magnitude. Thus, our setup is robust against the variation of the system size and our observations are not affected by finite-size effects. 

\section*{Variations of the Péclet Number}
\begin{figure}
	\centering
	\includegraphics[width=0.7\linewidth]{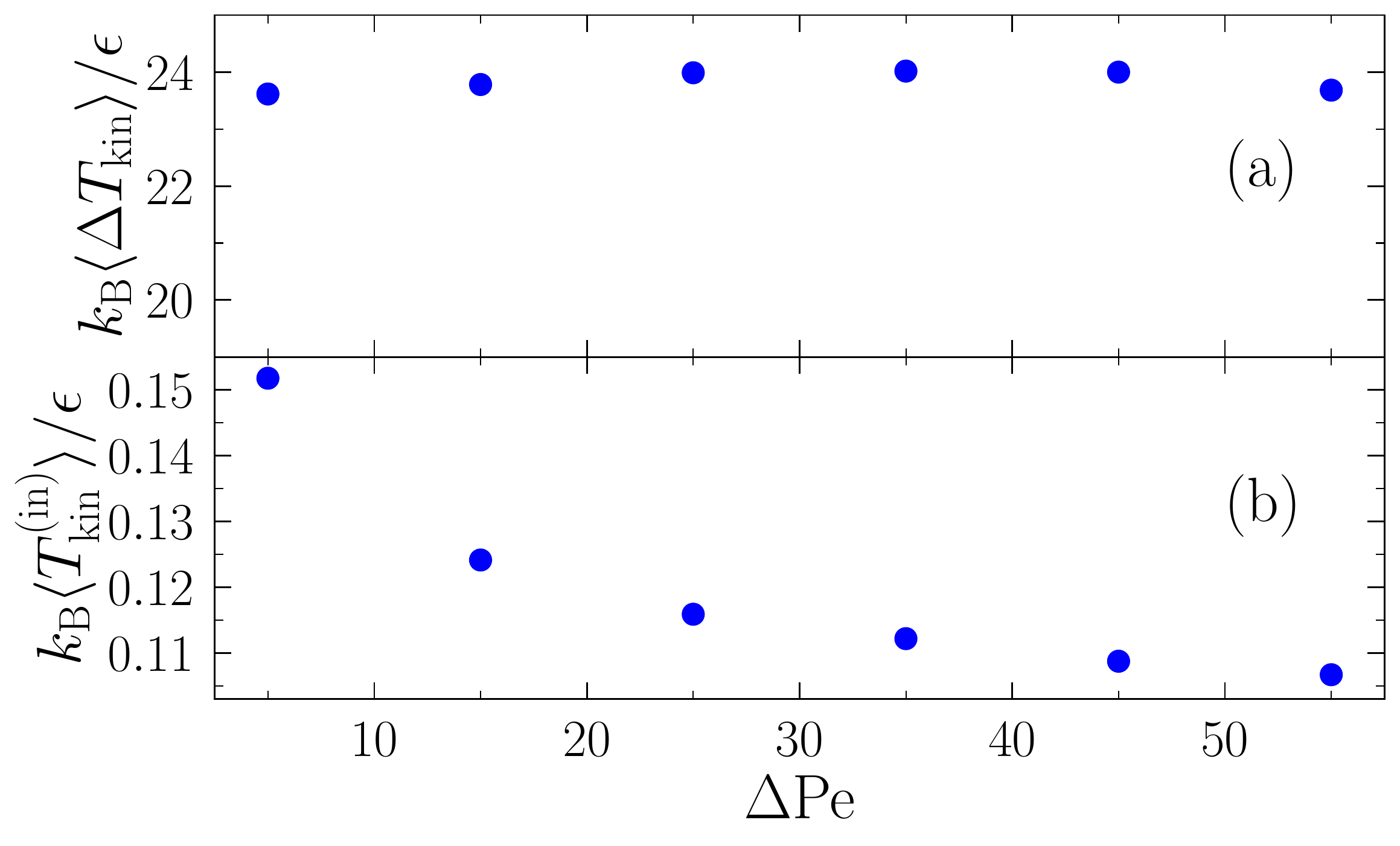}
	\caption{(a) Temperature difference $\Delta T_{\rm kin}=T_{\rm kin}^{\rm (out)}-T_{\rm kin}^{\rm (in)}$ between the refrigerator domain and its environment and (b) temperature in the refrigerator domain over $\Delta{\rm Pe}={\rm Pe}_{\rm out}-{\rm Pe}_{\rm in}$. All results are averaged over time in the steady state and over three realizations. Parameters: $N=16000$, $\varphi_{\rm tot}=0.35$, Pe$_{\rm in}\in\lbrace 55,65,75,85,95,105\rbrace$, Pe$_{\rm out}=110$, $m/(\gamma_{\text{t}}\tau_{\rm p})=5\times 10^{-2}$, $I/(\gamma_{\text{r}}\tau_{\rm p})=5\times10^{-3}$, $\epsilon/(k_{\text{B}}T_{\text{b}})=10$, $\sigma/\sqrt{D_{\rm r}D_{\rm t}}=1$, $x_0/L_x=0.1$.}
	\label{fig:figs8}
\end{figure}

As long as the requirements of regime (II) (see Fig.\ 1(b) of the main text) are met, the cooling effect is robust against variations of the choice of Péclet numbers. As we show in Fig.\ \ref{fig:figs8}(a), the (kinetic) temperature difference between the refrigerator domain and its environment is approximately invariant under variations of $\Delta{\rm Pe}={\rm Pe}_{\rm out}-{\rm Pe}_{\rm in}$ with Pe$_{\rm out}=110$ fixed and Pe$_{\rm in}$ varied. Remarkably, increasing $\Delta{\rm Pe}$ decreases the (kinetic) temperature in the refrigerator domain close to the lower limit of $\kb T_{\rm kin}/\epsilon=0.1$, which is given by the strength of the translational noise, as shown in Fig.\ \ref{fig:figs8}(b). Thus, the active refrigerator can be realized for very small differences in Péclet numbers but is still stable and even more efficient when the difference in Pe is increased. As a side remark, notice that even for choices of $\Delta{\rm Pe}$ (and $x_0$) which result in a left gray arrow in Fig.\ 1(b) of the main text [regime (II)] which is long enough to cross the upper transition line, we observe a significant cooling effect.

\section*{Role of Inertia}
\begin{figure}[h]
	\centering
	\includegraphics[width=0.7\linewidth]{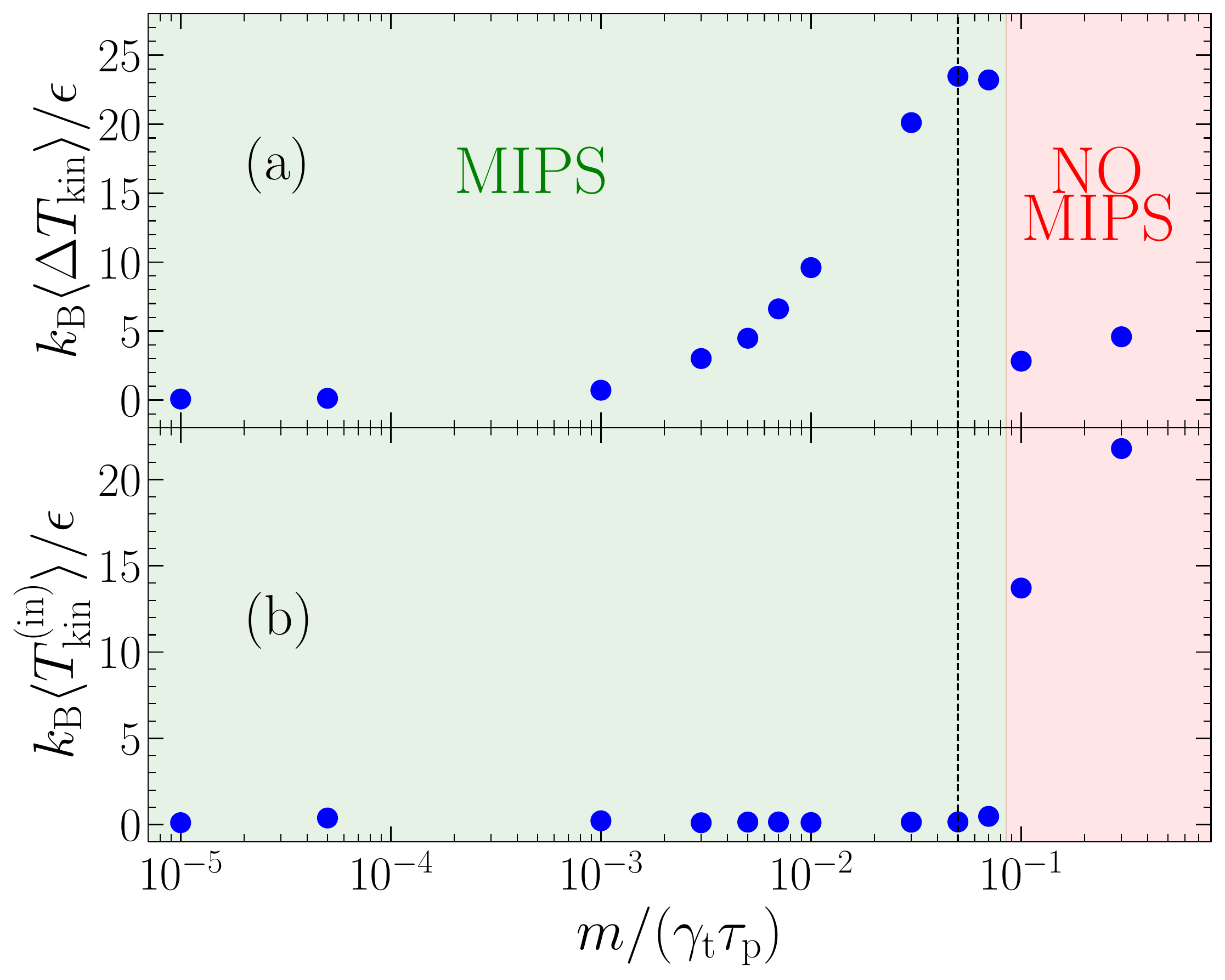}
	\caption{(a) Temperature difference $\Delta T_{\rm kin}=T_{\rm kin}^{\rm (out)}-T_{\rm kin}^{\rm (in)}$ between the refrigerator domain and its environment and (b) temperature in the refrigerator domain over the mass $m$ of the particles (averaged over time in the steady state and over three realizations). Motility-induced phase separations occurs in the green region while it is not possible in the red region (see also Ref.\ \cite{Mandal_PhysRevLett_2019} for a detailed discussion about the breakdown of MIPS at large inertia). The vertical dashed line indicates the value $m/(\gamma_{\rm t}\tau_{\rm p})=5\times 10^{-2}$ which we have used throughout this work. Parameters: $N=16000$, $\varphi_{\rm tot}=0.35$, Pe$_{\rm in}=105$, Pe$_{\rm out}=110$, $I/(\gamma_{\text{r}}\tau_{\rm p})=1/10\times m/(\gamma_{\rm t}\tau_{\rm p})$, $\epsilon/(k_{\text{B}}T_{\text{b}})=10$, $\sigma/\sqrt{D_{\rm r}D_{\rm t}}=1$.}
	\label{fig:figs9}
\end{figure}

In our study, we fixed the value of the mass $m/(\gamma_{\rm t}\tau_{\rm p})=0.05$. However, our results are valid even in a broader range of inertia as demonstrated in Fig.\ \ref{fig:figs9}: while for $m/(\gamma_{\rm t}\tau_{\rm p})\gtrsim 0.09$ motility-induced phase separation breaks down (see Ref.\ \cite{Mandal_PhysRevLett_2019} for a detailed discussion of the break down at large inertia), the temperature difference between the refrigerator domain and its environment decreases with decreasing inertia and finally vanishes when we are close to the overdamped regime at $m/(\gamma_{\rm t}\tau_{\rm p})=10^{-5}$. Although a temperature difference exists within the red region caused by the different Péclet numbers and different steady-state densities in the refrigerator domain and its environment, MIPS enhances the cooling effect significantly. Thus, both activity and inertia are crucial for the construction of an active refrigerator and a local maximum of the temperature difference can be observed for values of $m$ close to the breakdown of MIPS.

%\vspace{0.9cm}
\section*{Movies}
\emph{Movie M1 - active refrigerator:} The top panels show the profiles of the area fraction $\langle \varphi(x)\rangle$ and the temperature profiles $k_{\rm B}T_{\rm kin}(x)=m\langle |\vec{v}|^2\rangle_y/2$ averaged over the $y$ coordinate and 20 realizations with $N=16000$ particles for 
\begin{itemize}
	\item[(i)] uniform Pe (top left panels) with ${\rm Pe}_{\rm in}={\rm Pe}_{\rm out}=105$ and $\varphi_{\rm tot}=0.35$ (see also Fig.\ 2(a) of the main text) and 
	\item[(ii)] the active refrigerator (top right panels) with ${\rm Pe}_{\rm in}=105$, ${\rm Pe}_{\rm out}=110$, $\varphi_{\rm tot}=0.35$, and $x_0/L_x=0.1$ (see also Fig.\ 2(c) of the main text).
\end{itemize}
The lower panel shows an exemplary realization of the active refrigerator with the same parameters as in the top right panels but with $N=100,000$ particles (the corresponding profile of the area fraction $\langle \varphi(x)\rangle$ and the temperature profile $k_{\rm B}T_{\rm kin}(x)=m\langle |\vec{v}|^2\rangle_y/2$ averaged over the $y$ coordinate are shown as yellow dashed line in the top right panels). The other parameters are $m/(\gamma_{\text{t}}\tau_{\rm p})=5\times 10^{-2}$, $I/(\gamma_{\text{r}}\tau_{\rm p})=5\times10^{-3}$, $\epsilon/(k_{\text{B}}T_{\text{b}})=10$, and $\sigma/\sqrt{D_{\rm r}D_{\rm t}}=1$.

\vspace{0.2cm}
\emph{Movie M2 - absorbing and trapping tracers:} Example of the absorption and trapping of passive tracers (yellow) inside the active refrigerator with parameters ${\rm Pe}_{\rm in}=105$, ${\rm Pe}_{\rm out}=110$, $\varphi_{\rm tot}=0.35$, and $x_0/L_x=0.1$ (see also Fig.\ 5 of the main text). The passive tracers have the same attributes as the active particles (gray) except Pe=0 and are initially placed outside the targeted cooling domain. The fraction of passive tracers is given by $N_{\rm passive}/N=0.02$ with $N=16000$. All other parameters are the same as in Movie M1.

\end{document}